\documentclass[twocolumn]{aastex63}

\newcommand{\LD}[1]{{\color{black} #1}}
\newcommand{\BD}[1]{{\color{black} #1}}

\newcommand{\swift}{\emph{Swift}}
\newcommand{\nicer}{\emph{NICER}}

\newcommand{\fluxcgs}{erg~s$^{-1}$~cm$^{-2}$}
\newcommand{\lumcgs}{erg~s$^{-1}$}

\newcommand{\src}{Swift~J1818}

\accepted{August 27, 2020}

\begin{document}

\title{\nicer\ Observation of the Temporal and Spectral Evolution of Swift~J1818.0$-$1607: a Missing Link between Magnetars and Rotation Powered Pulsars}

\author[0000-0001-8551-2002]{Chin-Ping Hu}
\altaffiliation{JSPS International Research Fellow}
\affiliation{Extreme Natural Phenomena RIKEN Hakubi Research Team, \\RIKEN Cluster for Pioneering Research, 2-1 Hirosawa, Wako, Saitama 351-0198, Japan}
\affiliation{Department of Physics, National Changhua University of Education, Changhua 50007, Taiwan}
\affiliation{Department of Astronomy, Kyoto University, Kitashirakawa-Oiwake-cho, Sakyo-ku, Kyoto 606-8502, Japan}

\author[0000-0001-5072-8444]{Beste Begi\c{c}arslan}
\affiliation{Istanbul University, Science Faculty, Department of Astronomy and Space Sciences, Beyaz\i t, 34119, Istanbul, Turkey}

\author[0000-0002-3531-9842]{Tolga G\"uver}
\affiliation{Istanbul University, Science Faculty, Department of Astronomy and Space Sciences, Beyaz\i t, 34119, Istanbul, Turkey}
\affiliation{Istanbul University Observatory Research and Application Center, Istanbul University 34119, Istanbul Turkey}

\author[0000-0003-1244-3100]{Teruaki Enoto}
\affiliation{Extreme Natural Phenomena RIKEN Hakubi Research Team, \\RIKEN Cluster for Pioneering Research, 2-1 Hirosawa, Wako, Saitama 351-0198, Japan}

\author{George Younes}
\affiliation{Department of Physics, The George Washington University, Washington, DC 20052, USA}
\affiliation{Astronomy, Physics and Statistics Institute of Sciences (APSIS), The George Washington University, Washington, DC 20052, USA}

\author[0000-0001-6276-6616]{Takanori Sakamoto}
\affiliation{Department of Physics and Mathematics, Aoyama Gakuin University, 5-10-1 Fuchinobe, Chuo-ku, Sagamihara, Kanagawa 252-5258, Japan}

\author[0000-0002-5297-5278]{Paul S. Ray}
\affiliation{U.S. Naval Research Laboratory, Washington DC 20375, USA}

\author[0000-0001-7681-5845]{Tod E. Strohmayer}
\affiliation{Astrophysics Science Division and Joint Space-Science Institute, NASA's Goddard Space Flight Center, Greenbelt, MD 20771, USA}


\author{Sebastien Guillot}
\affiliation{IRAP, CNRS, 9 avenue du Colonel Roche, BP 44346, F-31028 Toulouse Cedex 4, France}

\author{Zaven Arzoumanian}
\affiliation{Astrophysics Science Division, NASA GSFC, 8800 Greenbelt Rd., Greenbelt, MD 20771, USA}

\author[0000-0001-7128-0802]{David M. Palmer}
\affiliation{Los Alamos National Laboratory, MS B244, PO Box 1663, Los Alamos, NM, 87545, USA}

\author{Keith C. Gendreau}
\affiliation{Astrophysics Science Division, NASA GSFC, 8800 Greenbelt Rd., Greenbelt, MD 20771, USA}

\author[0000-0002-0380-0041]{C.~Malacaria}
\altaffiliation{NASA Postdoctoral Fellow}
\affiliation{NASA Marshall Space Flight Center, NSSTC, 320 Sparkman Drive, Huntsville, AL 35805, USA}
\affiliation{Universities Space Research Association, Science and Technology Institute, 320 Sparkman Drive, Huntsville, AL 35805, USA}

\author[0000-0002-9249-0515]{Zorawar Wadiasingh}
\altaffiliation{NASA Postdoctoral Fellow}
\affiliation{Astrophysics Science Division, NASA GSFC, 8800 Greenbelt Rd., Greenbelt, MD 20771, USA}
\affiliation{Universities Space Research Association (USRA), Columbia, MD 21046, USA}

\author[0000-0002-6789-2723]{Gaurava K. Jaisawal}
\affil{National Space Institute, Technical University of Denmark, 
  Elektrovej 327-328, DK-2800 Lyngby, Denmark}

\author[0000-0002-4694-4221]{Walid A. Majid}
\affiliation{Jet Propulsion Laboratory, California Institute of Technology, Pasadena, CA 91109, USA}
\affiliation{California Institute of Technology, Pasadena, CA 91125, USA}

\begin{abstract}
\LD{We report on} the hard X-ray burst \BD{and the first $\sim$100 days \nicer\ monitoring of the soft X-ray temporal and spectral evolution} of the newly-discovered magnetar Swift~J1818.0$-$1607. The burst properties are typical of magnetars with a duration of $T_{90}=10\pm4$~ms and a temperature of $kT=8.4\pm0.7$~keV. \LD{The 2--8 keV pulse shows a broad, single peak profile with a pulse fraction increasing with time from 30\% to 43\%.} The \nicer\ observations reveal strong timing noise with $\dot{\nu}$ varying erratically by a factor of 10, with an average long-term spin-down rate of $\dot{\nu}=(-2.48\pm0.03)\times10^{-11}$~s$^{-2}$, implying \LD{an equatorial} surface magnetic field of $2.5\times10^{14}$~G and a young characteristic age of $\sim$470~yr. We detect a large spin-up glitch at MJD 58928.56 followed by a candidate spin-down glitch at MJD 58934.81, with no accompanying flux enhancements. The persistent soft X-ray spectrum of Swift~J1818.0$-$1607 can be modeled as an absorbed blackbody with a temperature of $\sim1$~keV. Its flux decayed by $\sim60$\% while the modeled emitting area decreased by $\sim30$\% over the \nicer\ observing campaign. This decrease, coupled with the increase in the pulse fraction points to a shrinking hot spot on the neutron star surface. \LD{Assuming a distance of 6.5 kpc, we measure a peak X-ray luminosity of $1.9\times10^{35}$~\lumcgs, lower than its spin-down luminosity of $7.2\times10^{35}$~\lumcgs. Its quiescent thermal luminosity is $\lesssim 1.7\times10^{34}$~\lumcgs, lower than those of canonical young magnetars.} We conclude that Swift~J1818.0$-$1607 is an important link between regular magnetars and high magnetic field rotation powered pulsars.

\end{abstract}
\keywords{pulsars: individual (Swift~J1818.0$-$1607), stars: neutron --- X-rays: stars --- stars: magnetars}


\section{Introduction} \label{sec:introduction}

Magnetars are a class of isolated neutron stars that manifest bright soft X-ray emission with $L_X\approx10^{31}$--$10^{36}$~\lumcgs\ and temperatures of $\sim0.5$~keV \LD{\citep[see, e.g.,][]{KaspiB2017, CotiZelatiRP2018}}. They occupy a unique place in the spin period versus the spin-down rate parameter space. Their long rotational periods of $P=2$--$12$~s and fast spin-down rates of $\dot{P}=10^{-13}$--$ 10^{-11}$~s~s$^{-1}$, imply high equatorial surface magnetic fields of $\sim10^{14}$~G and small characteristic ages ($\tau_c$) of, typically, a few thousand years \citep{KaspiB2017}. Given these temporal characteristics, the low inferred rotational energy losses cannot power magnetars' bright X-ray emission. Instead, they are believed to be powered by the decay of the extremely strong external and internal stellar magnetic fields \citep{Paczynski1992,DuncanT1992}.

Magnetars are highly variable X-ray sources. On short timescales they show hard X-ray bursts that last few hundred milliseconds. These can occur either in isolation or forming a ``storm'' with hundreds of bursts emitted within minutes to hours \citep{CollazziKH2015}. Recently, the magnetar SGR~1935+2154 entered such a burst active episode \citep{Palmer2020,YounesGE2020} and emitted a fast radio burst simultaneous to one of the X-ray bursts \citep{ScholzC2020, ZhangTX2020}. Following such bursting episodes, magnetars often undergo an outburst during which the persistent soft X-ray emission brightens by factors up to $\sim$1000 \citep{CotiZelatiRP2018}. \LD{During outbursts, their X-ray spectra often} show evidence of additional hotspots, of which the temperature and area decrease as the X-ray flux decays over time \citep{ReaIP2013, ZelatiRP2015}. Their spectral and temporal properties usually relax back to quiescence within months to years. We note that these magnetar-defining characteristics have been observed in other classes of neutron stars such as high magnetic field rotation powered pulsars (high-$B$ RPPs) \citep{GavriilGG2008, ArchibaldKT2016, GogusLK2016}, central compact objects \citep{ReaBE2016}, and low-$B$ magnetars \citep{ReaET2010}. Moreover, the canonical magnetar Swift~J1834.9$-$0846 shows a wind nebula, a trait of young RPPs \citep{YounesKK2016}. The observational evidence of high thermal luminosities of high-$B$ RPPs suggests that they may eventually exhibit magnetar-like behaviors \citep{KaspiM2005, NgK2011, HuNT2017}. Finally, four magnetars (PSR J1622$-$4950, PSR J1745$-$2900, XTE J1810$-$197, and 1E 1547.0$-$5408) have shown pulsed radio emission during outbursts, with properties that are usually different than those of RPPs and other magnetars \citep{CamiloRH2007, LevinBB2010, ShannonJ2013}. These results blur the boundary between the different classes of isolated neutron stars and perhaps hint at an evolutionary link among them \citep{ViganoRP2013}.

On 2020 March 12, the Burst Alert Telescope \citep[BAT;][]{BarthelmyBC2005} on board the \emph{Neil Gehrels Swift Observatory} \citep[hereafter \swift;][]{GehrelsCG2004} triggered an alert by a magnetar-like burst from a previously unknown source \citep[][the burst was also detected with Fermi-GBM, \citealt{MalacariaRV2020}]{EvansGK2020}, now named Swift~J1818.0$-$1607 (hereafter \src). A 1.36~s period was then discovered with the first follow-up observation with \nicer. This suggested \src\ is a new fast-spinning magnetar \citep{EnotoSY2020}. The periodicity was confirmed with radio observations and a period derivative of $\dot{P}=(8.16\pm0.02)\times10^{-11}$ s s$^{-1}$ was also reported. This initial timing solution implied an equatorial magnetic field of $3.4\times10^{14}$ G and $\tau_c=265$ yrs \citep{KaruppusamyDK2020,ChampionDJ2020}. The distance is estimated from the dispersion measure to be in the range of 4.8 to 8.1 kpc. In this study, we assume a distance of 6.5~kpc.


We report on \LD{(1)} the hard X-ray burst of \src\ with \swift\ BAT observations and \LD{(2) the} timing and spectral evolution \LD{of \src} with \nicer\ follow-up observations. We describe the \swift\ BAT and \nicer\ observation\LD{s} and data reduction in Section~\ref{sec:observation}. The properties of the hard X-ray burst are described in Section~\ref{sec:burst}. We introduce the timing analysis in Section~\ref{sec:timing}. The spectral properties and spectral evolution are described in Section~\ref{sec:spectrum}. Our interpretation of the observed phenomena is discussed in Section~\ref{sec:discussion} and summarized in Section~\ref{sec:summary}.

\section{Observation\LD{s}} \label{sec:observation}
\begin{table*}
    \caption{}
    \label{tab:bbodyrad}
    \begin{tabular}{cccccccc}
        \hline
        OBSID & Start~Time & Exposure & Count Rate & kT & Emitting Area  &  Flux$^{a}$ \\
        & (MJD) & (s.) & (cts/s) & (keV) & (km) &  \\ 
        \hline
        \hline

3201060101 & 58921.07 & 2536 & 1.60 $\pm 0.03$ & 1.04 $_{-0.02}^{+0.02}$ & 1.14 $_{-0.05}^{+0.05}$& 3.79 $_{-0.08}^{+0.08}$  \\
3556010101 & 58921.58 & 715  & 1.55 $\pm 0.05$ & 1.10 $_{-0.04}^{+0.05}$ & 1.03 $_{-0.08}^{+0.08}$& 3.86 $_{-0.13}^{+0.13}$  \\
3556010201 & 58922.10 & 536  & 1.49 $\pm 0.06$ & 1.05 $_{-0.04}^{+0.05}$ & 1.12 $_{-0.09}^{+0.10}$& 3.86 $_{-0.15}^{+0.15}$  \\
3556010202 & 58926.30 & 360  & 1.52 $\pm 0.07$ & 1.11 $_{-0.06}^{+0.06}$ & 1.03 $_{-0.11}^{+0.12}$& 3.95 $_{-0.18}^{+0.18}$  \\
3556010301 & 58927.40 & 1311 & 1.28 $\pm 0.04$ & 1.04 $_{-0.03}^{+0.03}$ & 1.05 $_{-0.07}^{+0.07}$& 3.21 $_{-0.10}^{+0.10}$  \\
3556010401 & 58928.37 & 2916 & 1.30 $\pm 0.02$ & 1.00 $_{-0.02}^{+0.02}$ & 1.14 $_{-0.05}^{+0.05}$& 3.21 $_{-0.07}^{+0.07}$  \\
3556010501 & 58929.06 & 4227 & 1.39 $\pm 0.02$ & 1.09 $_{-0.02}^{+0.02}$ & 0.99 $_{-0.04}^{+0.04}$& 3.43 $_{-0.06}^{+0.06}$  \\	      
3556010701 & 58930.22 & 3451 & 1.34 $\pm 0.02$ & 1.04 $_{-0.02}^{+0.02}$ & 1.06 $_{-0.04}^{+0.05}$& 3.31 $_{-0.06}^{+0.06}$  \\
3556010801 & 58931.46 & 1798 & 1.20 $\pm 0.03$ & 0.99 $_{-0.03}^{+0.03}$ & 1.11 $_{-0.06}^{+0.07}$& 3.00 $_{-0.08}^{+0.08}$  \\
3556010901 & 58932.09 & 3312 & 1.26 $\pm 0.02$ & 1.06 $_{-0.02}^{+0.02}$ & 1.00 $_{-0.04}^{+0.04}$& 3.11 $_{-0.06}^{+0.06}$  \\   
3556011001 & 58933.83 & 1449 & 1.21 $\pm 0.03$ & 1.03 $_{-0.03}^{+0.03}$ & 1.02 $_{-0.06}^{+0.07}$& 2.99 $_{-0.08}^{+0.08}$  \\
3556011101 & 58934.34 & 2888 & 1.20 $\pm 0.02$ & 1.03 $_{-0.02}^{+0.02}$ & 1.03 $_{-0.05}^{+0.05}$& 2.98 $_{-0.06}^{+0.06}$  \\	      
3556011201 & 58935.84 & 2414 & 1.17 $\pm 0.03$ & 1.11 $_{-0.03}^{+0.03}$ & 0.88 $_{-0.05}^{+0.05}$& 2.88 $_{-0.07}^{+0.07}$  \\
3556011301 & 58936.67 & 2410 & 1.17 $\pm 0.03$ & 1.05 $_{-0.03}^{+0.03}$ & 0.98 $_{-0.05}^{+0.05}$& 2.89 $_{-0.07}^{+0.07}$  \\
3556011401 & 58937.58 & 2786 & 1.11 $\pm 0.02$ & 1.02 $_{-0.02}^{+0.02}$ & 0.99 $_{-0.05}^{+0.05}$& 2.74 $_{-0.06}^{+0.06}$  \\
3556011501 & 58938.35 & 1223 & 0.97 $\pm 0.03$ & 1.09 $_{-0.05}^{+0.05}$ & 0.83 $_{-0.07}^{+0.07}$& 2.43 $_{-0.09}^{+0.09}$  \\
3556011502 & 58939.52 & 2251 & 1.18 $\pm 0.03$ & 1.04 $_{-0.03}^{+0.03}$ & 0.98 $_{-0.05}^{+0.05}$& 2.90 $_{-0.07}^{+0.07}$  \\	      
3556011503 & 58940.56 & 2373 & 1.14 $\pm 0.03$ & 1.00 $_{-0.03}^{+0.03}$ & 1.05 $_{-0.06}^{+0.06}$& 2.79 $_{-0.07}^{+0.07}$  \\
3556011601 & 58942.49 & 3449 & 1.20 $\pm 0.02$ & 1.03 $_{-0.02}^{+0.02}$ & 1.03 $_{-0.04}^{+0.05}$& 2.93 $_{-0.06}^{+0.06}$  \\
3556011701 & 58944.57 & 2506 & 1.04 $\pm 0.02$ & 1.03 $_{-0.03}^{+0.03}$ & 0.96 $_{-0.05}^{+0.05}$& 2.60 $_{-0.06}^{+0.06}$  \\
3556011801 & 58947.21 & 2933 & 1.05 $\pm 0.02$ & 1.06 $_{-0.03}^{+0.03}$ & 0.90 $_{-0.04}^{+0.05}$& 2.58 $_{-0.06}^{+0.06}$  \\
3556012001 & 58953.86 & 1888 & 1.06 $\pm 0.03$ & 0.97 $_{-0.03}^{+0.03}$ & 1.07 $_{-0.07}^{+0.07}$& 2.59 $_{-0.07}^{+0.07}$  \\
3556012101 & 58953.92 & 1708 & 0.99 $\pm 0.03$ & 1.05 $_{-0.04}^{+0.04}$ & 0.89 $_{-0.06}^{+0.07}$& 2.42 $_{-0.07}^{+0.08}$  \\
3556012102 & 58956.06 & 734.1& 0.80 $\pm 0.05$ & 1.10 $_{-0.07}^{+0.08}$ & 0.76 $_{-0.09}^{+0.10}$& 2.06 $_{-0.12}^{+0.12}$  \\
3556012201 & 58959.80 & 1742 & 1.30 $\pm 0.03$ & 1.12 $_{-0.04}^{+0.04}$ & 0.90 $_{-0.06}^{+0.06}$& 3.16 $_{-0.09}^{+0.09}$  \\
3556012301 & 58962.32 & 3552 & 0.86 $\pm 0.02$ & 1.07 $_{-0.03}^{+0.03}$ & 0.80 $_{-0.04}^{+0.04}$& 2.12 $_{-0.05}^{+0.05}$  \\
3556012401 & 58966.32 & 2362 & 0.94 $\pm 0.02$ & 1.08 $_{-0.03}^{+0.03}$ & 0.83 $_{-0.05}^{+0.05}$& 2.30 $_{-0.06}^{+0.06}$  \\
3556012501 & 58968.91 & 2148 & 0.85 $\pm 0.02$ & 1.11 $_{-0.04}^{+0.04}$ & 0.75 $_{-0.05}^{+0.05}$& 2.13 $_{-0.06}^{+0.06}$  \\
3556012601 & 58971.36 & 3580 & 0.91 $\pm 0.02$ & 1.09 $_{-0.03}^{+0.03}$ & 0.80 $_{-0.04}^{+0.04}$& 2.25 $_{-0.05}^{+0.05}$  \\
3556012602 & 58972.26 & 3978 & 0.85 $\pm 0.02$ & 1.10 $_{-0.03}^{+0.03}$ & 0.77 $_{-0.04}^{+0.04}$& 2.11 $_{-0.05}^{+0.05}$  \\
3556012701 & 58974.00 & 1839 & 0.85 $\pm 0.03$ & 1.04 $_{-0.04}^{+0.04}$ & 0.84 $_{-0.06}^{+0.06}$& 2.10 $_{-0.07}^{+0.07}$  \\
3556012801 & 58987.18 & 809.1& 0.79 $\pm 0.04$ & 0.99 $_{-0.05}^{+0.06}$ & 0.90 $_{-0.10}^{+0.11}$& 2.00 $_{-0.09}^{+0.10}$  \\
3556012901 & 58989.89 & 3653 & 0.71 $\pm 0.02$ & 1.06 $_{-0.04}^{+0.05}$ & 0.74 $_{-0.04}^{+0.04}$& 1.75 $_{-0.04}^{+0.04}$  \\
3556013001 & 58993.13 & 1741 & 0.81 $\pm 0.03$ & 1.08 $_{-0.05}^{+0.05}$ & 0.75 $_{-0.06}^{+0.07}$& 1.93 $_{-0.07}^{+0.07}$  \\
3556013201 & 58998.99 & 783  & 0.67 $\pm 0.03$ & 1.05 $_{-0.07}^{+0.07}$ & 0.76 $_{-0.09}^{+0.10}$& 1.76 $_{-0.09}^{+0.09}$  \\
3556013202 & 58999.05 & 2346 & 0.80 $\pm 0.02$ & 1.01 $_{-0.03}^{+0.03}$ & 0.87 $_{-0.05}^{+0.06}$& 1.98 $_{-0.06}^{+0.06}$  \\
3556013301 & 59001.96 & 1825 & 1.06 $\pm 0.03$ & 1.05 $_{-0.04}^{+0.04}$ & 0.91 $_{-0.06}^{+0.06}$& 2.55 $_{-0.07}^{+0.07}$  \\
3556013401 & 59004.99 & 1922 & 0.63 $\pm 0.02$ & 1.01 $_{-0.04}^{+0.04}$ & 0.77 $_{-0.06}^{+0.07}$& 1.57 $_{-0.06}^{+0.06}$  \\
3556013501 & 59008.88 & 2350 & 0.68 $\pm 0.02$ & 1.01 $_{-0.04}^{+0.04}$ & 0.80 $_{-0.06}^{+0.06}$& 1.68 $_{-0.05}^{+0.05}$  \\
3556013502 & 59009.01 & 1063 & 0.75 $\pm 0.03$ & 1.02 $_{-0.05}^{+0.05}$ & 0.82 $_{-0.08}^{+0.09}$& 1.85 $_{-0.08}^{+0.08}$  \\
3556013601 & 59011.20 & 1812 & 0.48 $\pm 0.02$ & 1.02 $_{-0.05}^{+0.06}$ & 0.67 $_{-0.07}^{+0.08}$& 1.22 $_{-0.06}^{+0.06}$  \\
3556013701 & 59014.09 & 1530 & 0.85 $\pm 0.03$ & 1.37 $_{-0.07}^{+0.08}$ & 0.52 $_{-0.04}^{+0.05}$& 2.21 $_{-0.09}^{+0.09}$  \\
3556013801 & 59017.60 & 4006 & 0.73 $\pm 0.02$ & 1.07 $_{-0.03}^{+0.03}$ & 0.74 $_{-0.04}^{+0.04}$& 1.79 $_{-0.04}^{+0.04}$  \\
3556013901 & 59020.42 & 4011 & 0.85 $\pm 0.02$ & 1.09 $_{-0.03}^{+0.03}$ & 0.76 $_{-0.04}^{+0.04}$& 2.07 $_{-0.05}^{+0.05}$  \\
3556014001 & 59023.27 & 1275 & 0.59 $\pm 0.03$ & 0.97 $_{-0.05}^{+0.05}$ & 0.82 $_{-0.08}^{+0.09}$& 1.49 $_{-0.07}^{+0.07}$  \\
3556014301 & 59023.33 &	1875 & 0.50 $\pm 0.02$ & 1.03 $_{-0.05}^{+0.05}$ & 0.67 $_{-0.06}^{+0.07}$& 1.27 $_{-0.06}^{+0.06}$  \\
         \hline
\tabularnewline
\end{tabular}
\\ {\footnotesize{
$^{a}$ 0.3$-$10 keV flux values are unabsorbed and in units of $10^{-11}$ \fluxcgs.}
}
    \end{table*}

\subsection{\swift\ BAT}
The BAT triggered an alert at 21:16:47.328 UTC on 2020 March 12 for a short burst located at R.A.$=18^h\,18^m\,00.2^s$ and Decl.$=-16^{\circ}\,07\arcmin\,52.3\arcsec$, which was refined with prompt observation with X-Ray Telescope onboard \swift. The on-board trigger occurred on an 8 ms \LD{timescale} in the 15--50 keV band with a rate significance of 24$\sigma$.  The on-board calculated location was R.A.$=$18h~17m~53s and Decl.$=-16^{\circ}$~06\arcmin~05\arcsec\ \citep{EvansGK2020}. 
HEAsoft version 6.27 \citep{Heasarc2014} and \swift\ BAT CALDB (version 20171016) were used for the BAT data analysis.  We use the raw counts (non-mask-weighted) data for the temporal analysis to maximize the signal-to-noise ratio of the light curve.  For spectral analysis, the standard mask-weighted analysis is performed.  


\subsection{\nicer}
\nicer\ is a non-imaging soft X-ray telescope onboard the \emph{International Space Station}. \LD{It has absolute timing uncertainty better than $300$~ns}. After the \swift\ BAT detection of the short burst, a series of follow-up observations was began at 01:38 UTC on 2020 March 13 with \nicer. Through 2020 \LD{June 23}, we have monitored this source with \nicer\ for a total exposure of \LD{$\sim102$} ks. All observations used for the present analysis are listed in Table \ref{tab:bbodyrad}. The basic data processing was carried out with NICERDAS version 7 in HEAsoft 6.27.2 and \nicer\ calibration database version 20200202. We created cleaned event files by applying the standard calibration and filtering tool \texttt{nicerl2} to the unfiltered data. We performed barycentric correction using \texttt{barycorr} with the JPL solar system ephemeris DE405 and the refined source position \citep{EvansGK2020}. 


\begin{figure}
\includegraphics[width=0.47\textwidth]{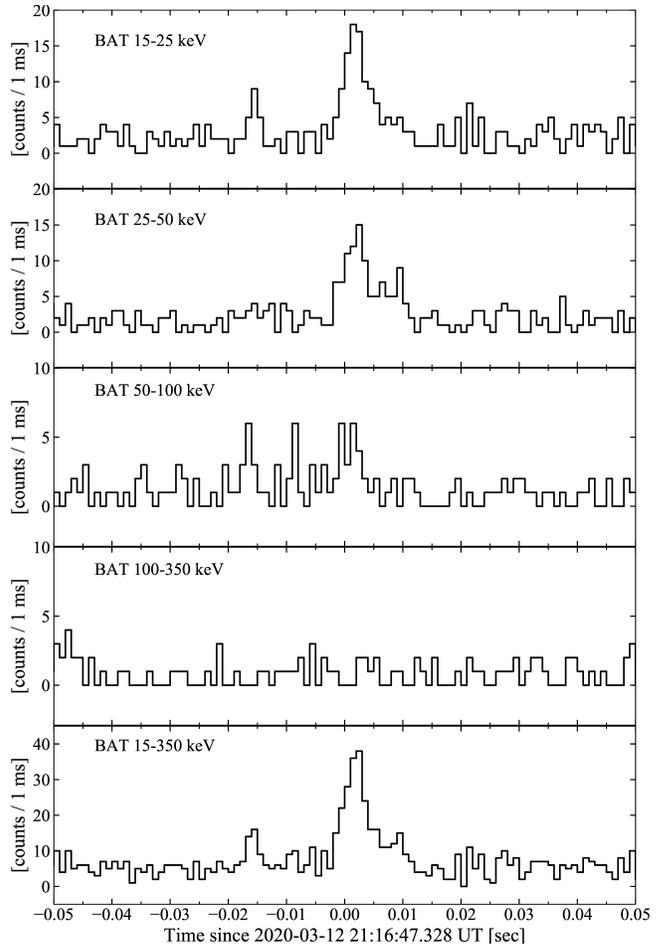} 
\caption{\LD{The \swift\ BAT light curves, in 1 ms bins and five energy bands (15--25, 25--50, 50--100, 100--350, 15--350 keV), around the time of the first short burst of \src.} \label{fig:bat_lc}}
\end{figure}

\section{Data Analysis and Result} 

\subsection{Hard X-ray Burst}\label{sec:burst}
We created non-mask-weighted light curves around the BAT trigger time (Figure \ref{fig:bat_lc}). No significant emission above 100 keV is seen by BAT, which confirms the soft nature of the burst. A clear pulse with a fast rise and an exponentially decaying tail is observed.  We used the \texttt{battblocks} tool to estimate the T$_{\rm{90}}$ duration, which is defined as the time interval where the integrated photon counts increase from 5\% to 95\% of the total counts, as $10\pm4$ ms (15--350 keV). The burst profile can be well fit with the QDP\footnote{https://heasarc.gsfc.nasa.gov/ftools/others/qdp/qdp.html} BURS model \LD{(a linear rise followed by an exponential decay)}, where the rising duration is $4.6\pm0.4$~ms and the \LD{e-folding time} of the decay is $2.4\pm0.5$~ms. The \LD{T$_{\rm{90}}$ duration} of this burst is on the short end of typical short bursts of SGRs. \LD{We note that a burst-like feature can be marginally seen at $T_{\rm{burst}}-0.015$~s (\LD{where T$_{\rm{burst}}$ is the time of the peak}) but it is likely an instrumental effect.} 

\begin{figure*}
\begin{minipage}{0.49\linewidth}
\includegraphics[width=0.99\textwidth]{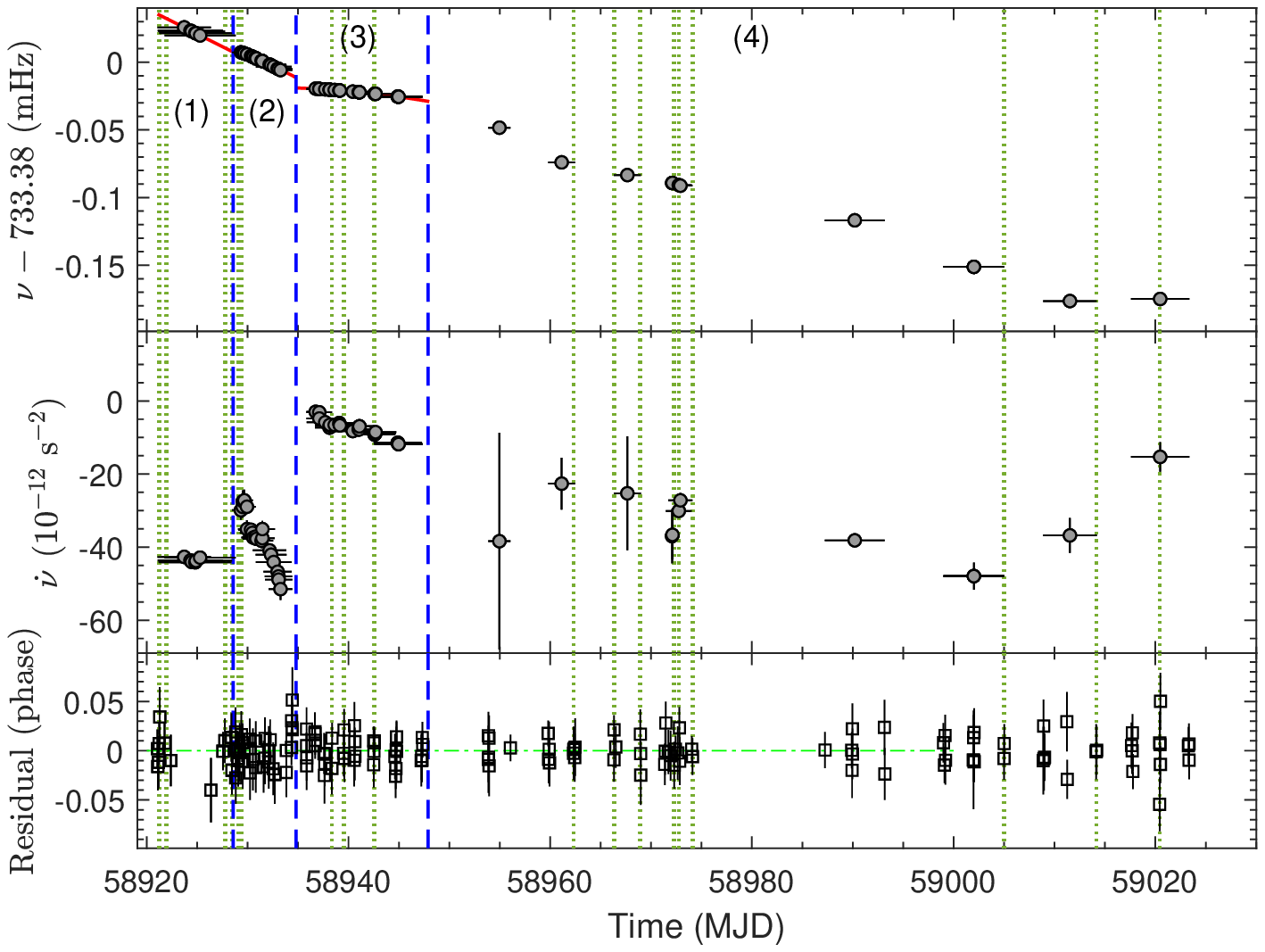} 
\end{minipage}
\hspace{0.0cm}
\begin{minipage}{0.49\linewidth}
\includegraphics[width=0.99\textwidth]{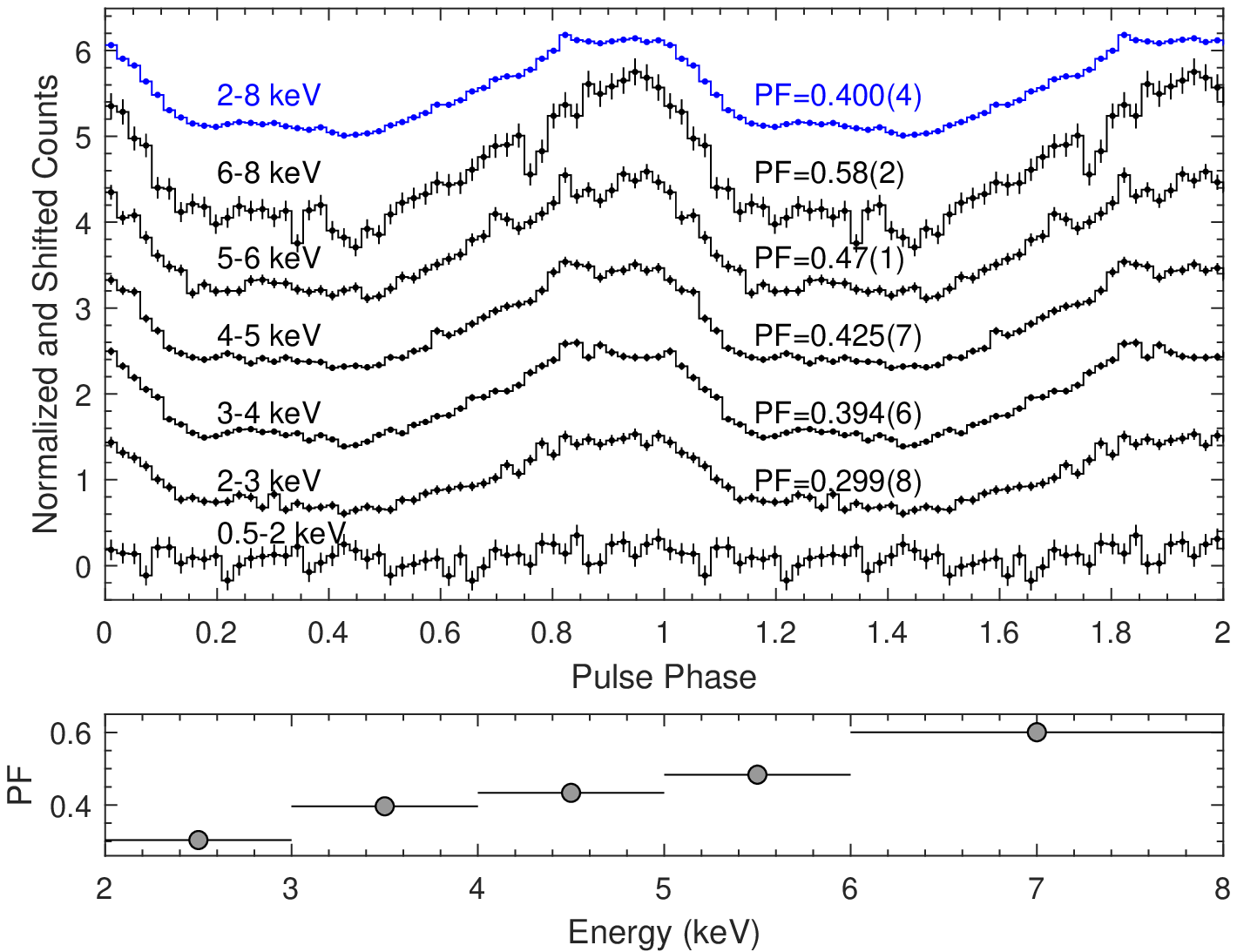}
\end{minipage}
\caption{(Left) Evolution of the timing properties of \src. The upper panel shows the pulse frequency evolution, while the red lines show the best timing solutions for three segments. The evolution of $\dot{\nu}$ is shown in the middle panel. The phase residuals are shown in the bottom panel. The blue dashed vertical lines indicate the boundaries of four segments, while the first two are \LD{timing discontinuities}. Green dotted lines denote the times of short soft-band X-ray bursts seen with \nicer. (Right) Pulse profile of \src\ in 0.5--2 keV, 2--3 keV, 3--4 keV, 4--5 keV, 5--6 keV, 6--8 keV, and 2--8 keV obtained from all the \nicer\ observations listed in Table \ref{tab:bbodyrad}. The bottom panel shows the energy dependence of the pulse fraction. \label{fig:phase_residual_profile}}
\end{figure*}

We then extracted the BAT spectrum of this burst from the interval between $T_{\rm{burst}}-2$~ms and $T_{\rm{burst}}+12$~ms (the T$_{\rm{100}}$ duration). Because of the low statistics of the spectrum, we generated a 16 channel spectrum instead of the standard 80 channels with \texttt{batbinevt}. We fit the spectrum with three models, including a blackbody, a bremsstrahlung, and a power law. The spectral analysis was performed with \texttt{XSPEC} v12.11.0. The best fit model is a blackbody with statistics of $\chi^2/\mathrm{dof}=5.67/11$ \LD{where dof denotes the degrees of freedom}.  The $\chi^2/\mathrm{dof}$ of the bremsstrahlung and the power-law models are 13.4/11 and 19.9/11, respectively. The temperature of the blackbody model is $8.4\pm0.7$~keV with a radius of $2.4\pm0.5$~km. The average 15--150 keV flux is $(6.1\pm0.8)\times10^{-7}$ \fluxcgs\ and the fluence of this burst \LD{is} ($8.5\pm1.1)\times10^{-9}$~erg~cm$^{-2}$. The isotropic equivalent luminosity is $\sim3.1\times10^{39}$~\lumcgs\ and for a total energy of $\sim4.3\times10^{37}$~erg. The uncertainties in this paper denote the 68\% confidence intervals unless stated otherwise.


\subsection{Spin Period Evolution and Soft X-ray Bursts} \label{sec:timing}
We selected \nicer\ events in the energy range of 2--8 keV for timing analysis. The average count rate between 2020 March 13 and June 23 was $\sim1.1$~counts~s$^{-1}$. We first found a coherent pulsation with a frequency of 0.733417(4) Hz ($P\approx1.36$~s) and a single-peaked pulse profile in the 2020 March 13 data set \citep[ObsID 3201060101,][]{EnotoSY2020}. Due to limited visibility, \src\ could not be observed with NICER from March 15 – 17. Then \nicer\ performed a series of monitoring observations with a cadence of roughly one day until MJD 58948 (2020 April 9). This allowed us to track the spin period evolution through a phase-coherent analysis. After MJD 58953 (2020 April 14), we revised the cadence down to $\sim$ 2--5 days, \LD{resulting in ambiguities of cycle counts during large gaps. Therefore, we did not perform the phase-coherent analysis afterward. }

\begin{deluxetable*}{lccccc}
\tablecaption{Spin parameters of \src. We also derive the physical quantities of characteritic age ($\tau_c$), dipolar magnetic field ($B$), and the spin-down luminosity ($L_{\rm{sd}}$) based on the long-term average $\nu$ and $\dot{\nu}$.\label{tab:ephemeris}} 
\tablehead{\colhead {Parameter} & \colhead {Segment 1} & \colhead{Segment 2} & \colhead{Segment 3} & \colhead{Segment 4} & \colhead{Long-Term} }
\startdata
MJD Start & 58921.07 & 58928.81 & 58935.83 & 58953.86 & 58921.07 \\
MJD End & 58928.45 & 58933.84 & 58947.35 & 59023.32 &  59023.32  \\
$T_0$ (MJD) & 58922.31 & 58930.751766 & 58940.301565 & 58987.72 & 58972.10 \\
$\nu$ (Hz) & 0.7334109(1) & 0.73338323(5) & 0.73335840(1) & 0.7332600(1) & 0.7332929(1) \\
$\dot{\nu}$ ($10^{-12}$ s$^{-2}$) & $-43.2(6)$ & $-37.3(7)$ & $-8.88(5)$ & $-22.8(1)$ &  $-24.8(3)$\\
$\ddot{\nu}$ ($10^{-18}$ s$^{-3}$) & \nodata\ & $-38(9)$ & $-8.9(6)$ & \nodata\ & \nodata \\
rms residual (phase) & 0.015 & 0.018 &  0.012 & \nodata\ & \nodata\ \\
$\tau_c$ (yr) & 270 & 310 & 1300 & 510 & 470 \\
$B$ ($10^{14}$ G) & $3.4$ & $3.1$ & $1.5$ & 2.4 & $2.5$ \\
$L_{\rm{sd}}$ ($10^{35}$ \lumcgs) & $13$ & $11$ & $2.6$ & 6.6 & $7.2$ 
\enddata
\end{deluxetable*}

\LD{We used the pulse profile smoothed by the local polynomial regression method \citep{ClevelandL1996} as the template to calculate phase shifts and corresponding times of pulse arrival (TOAs) for the phase-coherent analysis. \LD{In each observation, we divided the time into a few segments such that each contained 1000 photons,} and calculated TOAs using the maximum likelihood analysis method described in \citet{LivingstoneRC2009}. } We found that the spin frequency $\nu$ and the spin-down rate $\dot{\nu}$ of \src\ is highly variable. At least two timing anomalies in the first month of \nicer\ monitoring were observed. The first one, consistent with a traditional spin-up glitch, occurred at MJD 58928.56 (2020 March 20) with a size of $\Delta\nu=(2.7\pm0.1)\times10^{-6}$~Hz ($\Delta\nu/\nu=(3.7\pm0.2)\times10^{-6}$) and $\Delta\dot{\nu}=(5.1\pm0.5)\times10^{-12}$~s$^{-2}$ ($\Delta\dot{\nu}/\dot{\nu}=-0.12\pm0.01$). The second one can be described as an anti-glitch with $\Delta\nu=(-5.28\pm0.01)\times10^{-6}$~Hz ($\Delta\nu/\nu=(-7.20\pm0.01)\times10^{-6}$) and $\Delta\dot{\nu}=(4.69\pm0.05)\times10^{-11}$~s$^{-2}$ ($\Delta\dot{\nu}/\dot{\nu}=-0.91\pm0.01$) at MJD 58934.81 (2020 March 26). However, we could not rule out the possibility that the frequency evolves continuously and dramatically instead of an abrupt jump \LD{due to a gap in coverage at the epoch of the timing anomaly}. We divided the observations before MJD 58948 (2020 April 9) into three segments according to these two timing discontinuities and fit TOAs with second-order or third-order polynomials individually. The timing solutions for individual segments are summarized in Table \ref{tab:ephemeris}.

To obtain the evolution of $\nu$ and $\dot{\nu}$, we used the technique described in \cite{DibK2014} by choosing a window that contains 8--15 consecutive TOAs and fitting their phases with a second-order polynomial. Similar to the moving average technique, we moved the window with a step adaptively equal to a separation of 2--4 consecutive TOAs over the entire segments. For data beyond MJD \LD{58948 (2020 April 9)}, which is noted as segment 4, \BD{we did not perform phase-coherent analysis spanning the entire segment due to the ambiguity of cycle counts in a few large gaps}. The result is shown in Figure \ref{fig:phase_residual_profile}. Since the onset of the outburst, the source shows a high level of timing noise, in which $\dot{\nu}$ significantly changes on a timescale of a few days. We derived a long-term \LD{$\dot{\nu}=(-2.48\pm0.03)\times10^{-11}$~s$^{-2}$ ($\dot{P}=(4.61\pm0.06)\times 10^{-11}$~s~s$^{-1}$)} by fitting \LD{the spin frequency evolution over the entire time span with a first-order polynomial function}. This results in a characteristic age of \LD{$\tau_c=470$~yr} by assuming a braking index of 3 and rapid spin at birth. The surface equatorial magnetic field can be inferred as \LD{$B=2.5\times10^{14}$~G} and the $L_{\rm{sd}}$ is estimated as \LD{$7.2\times10^{35}$~\lumcgs}. However, the dramatic changes in the timing behavior make it difficult to conclusively characterize the long-term timing properties at the current stage. If we consider the timing solution in individual epochs, the derived $\tau_c$ could be in a wide range of 270--1300 yr, and $L_{\rm{sd}}$ could be in the range of $(2.6$--$13)\times10^{35}$~\lumcgs.

\begin{figure*}
    \begin{minipage}{0.48\linewidth}
    \includegraphics[width=0.99\textwidth]{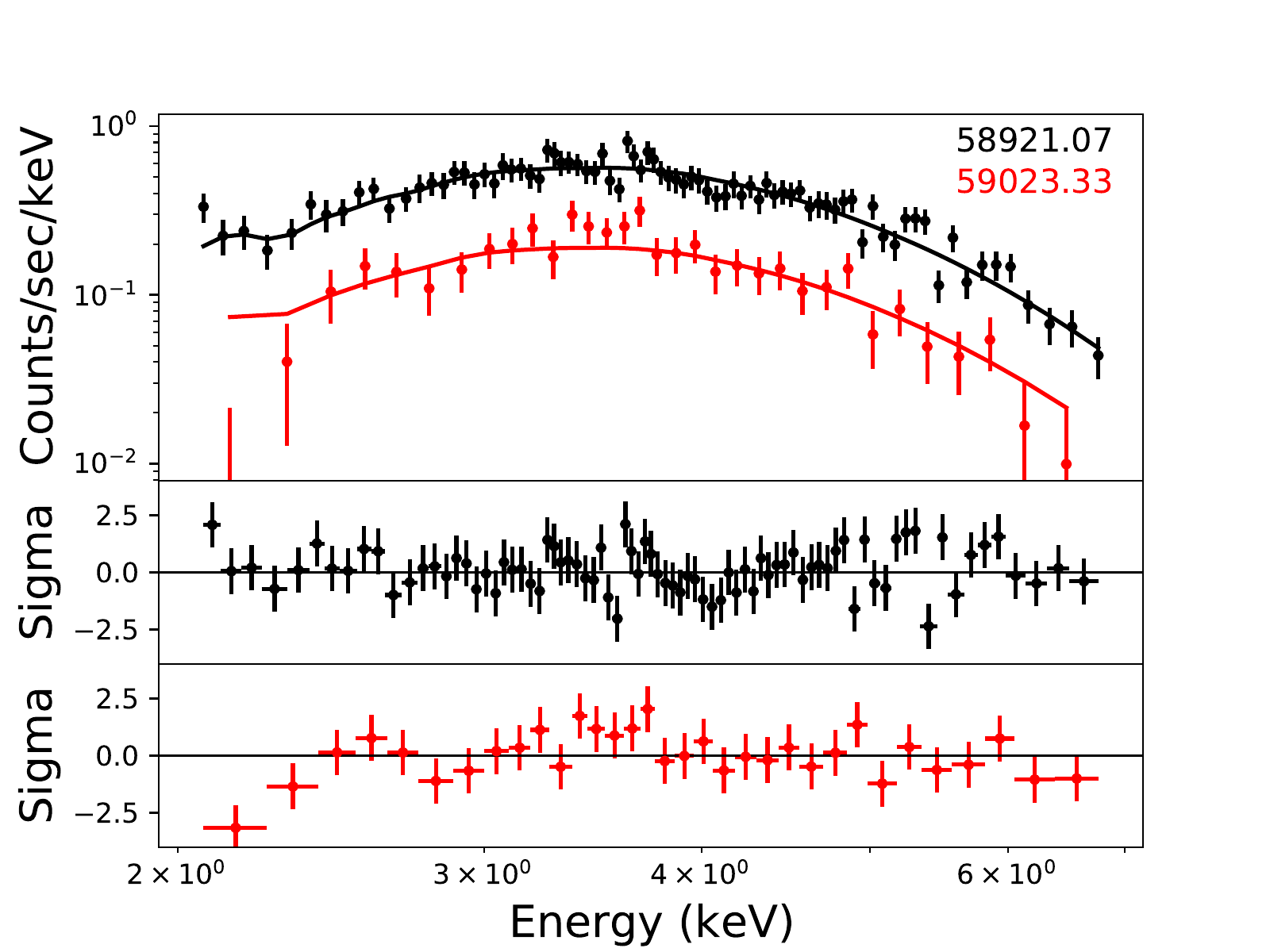} 
    \end{minipage}
    \hspace{0.0cm}
    \begin{minipage}{0.48\linewidth}
    \includegraphics[width=0.95\textwidth]{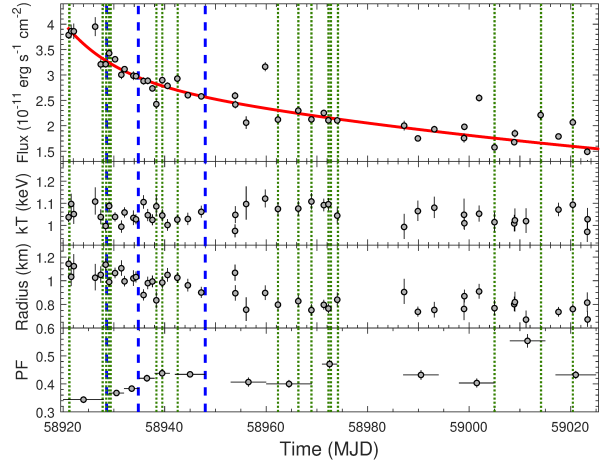} 
    \end{minipage}
    \caption{(left) \nicer~ spectra obtained from the first (black) and the last (red) observations together with the best fit absorbed blackbody models. 
    \LD{The dates of the observations are indicated.} Lower panels show the residuals from the blackbody models. (right) Spectral evolution of \src; from upper to lower, panels show the evolution of the unabsorbed flux, temperature, apparent emitting radius, and the pulse fraction in 2--8 keV. Vertical green dashed lines show the burst times, while the vertical blue dashed lines show the times of the timing anomalies. The red solid curve shows the best-fit double exponential decay trend. Unabsorbed flux values are calculated for the 0.3$-$10~keV band.}
    \label{fig:spdata}
\end{figure*}

To see whether the pulse profile is energy-dependent, we created energy-resolved pulse profiles in six bands: 0.5--2, 2--3, 3--4, 4--5, 5--6, and 6--8 keV (see the right panel of Figure \ref{fig:phase_residual_profile}). The pulsation cannot be seen below 2 keV due to heavy interstellar absorption. The 2--8~keV folded light curve shows a broad asymmetric peak. No significant energy dependence of the pulse shape is seen across 2--8~keV. We calculated the rms pulse fraction \citep[PF, see definition in][]{DibKG2009, AnAH2015} in these energy bands and found that the PF increases with energy. The background estimated from \texttt{nibackgen3C50}{\footnote{https://heasarc.gsfc.nasa.gov/docs/nicer/tools/nicer\_bkg\_est\_tools.html}} is subtracted from the pulse profile. To test whether the pulse profile changes following glitches, we created time-resolved pulse profiles. We divided the X-ray \LD{observations} into \LD{thirteen} time bins \LD{to ensure each one has effective exposure time $\gtrsim5000$ s}: MJD 58920--58928, 58929--58932, 58932--58935, 58935--58938, 58938--58941, 58942--58948, 58953--58960, 58962--58969, 58971--58974, 58987--58994, 58998--59005, 59008--59015, and 59017--59024. We did not \LD{observe} any significant changes in the shape of the pulse profile accompanying either timing discontinuity. The PF increased from 0.34(1) to $\sim0.43$ in the first $\sim15$ days from the onset of the outburst and fluctuated \LD{around 0.43 except for an extreme value during MJD 59008--59015 (see Figure \ref{fig:spdata})}. \LD{We found that the background level estimated with \texttt{nibackgen3C50} is extremely high on MJD 59014. This could result in an over-estimate of the PF \BD{if the background in this observation is not accurately estimated}.}

In addition to tracking the evolution of the timing behavior and the variability of the pulse profile, we searched for SGR-like X-ray bursts. We created light curves with a bin size of 1/256 s. For each time bin, we calculated the mean count rate in the surrounding 20 seconds. The \LD{probability of the photon distribution in each time bin can} be evaluated with the Poisson distribution. During the \nicer\ campaign, we identified \LD{21} bursts with a detection significance higher than 5$\sigma$. \LD{We also emplyed other time bin sizes to reconfirm the detection of bursts.} Their occurrence times are indicated in Figure \ref{fig:phase_residual_profile}. Four of them are clustered near the first glitch and three of them are near MJD 58972. The candidate anti-glitch likely occurred during the observational gap that prevented us from probing the association between \LD{radiative events} and the anti-glitch. We noticed that the light curve observed in the last GTI of MJD 58944 (ObsID 3556011701) exhibited many burst-like structures with a \LD{timescale much longer than that of regular bursts and} accompanied an enhancement of the baseline level \LD{with a duration of} $\sim300$ s. After carefully examining the photon distribution in different detectors and the timing/spectral behavior during this period we suggest it was caused by a particle flaring episode, and was not intrinsic to \src.

\subsection{Spectral Analysis} \label{sec:spectrum}

To monitor the long-term spectral evolution of \src\ with \nicer\ we extracted X-ray spectra from the burst-free times using \texttt{XSELECT}. We grouped each spectrum to have 50 counts per channel using \texttt{grppha} and used \texttt{XSPEC} version 12.10.0c \citep{Arnaud1996} to fit the spectra. We created background files for each observation using the \texttt{nibackgen3C50} tool. \LD{We used the response and ancillary response files currently in the \nicer\ calibration database. Note that we removed the data from Focal Plane Modules 14 and 34 and used an adjusted ancillary response file accordingly.}


The very large column density along the line of sight to \src\ causes background counts to dominate below 2~keV. Also, above 7~keV the signal-to-noise ratio of the source decreases significantly, therefore we only perform our fits in the 2$-$7~keV band. We modeled the X-ray spectra with an absorbed blackbody model. To determine the amount of Hydrogen column density ($N_{\rm{H}}$), we used the \texttt{tbabs} model with ISM abundances \citep{WilmsAM2000}.


\LD {To model the spectral evolution,} we allowed all the parameters of the models of individual data sets to be free except for the $N_{\rm{H}}$ which was kept linked. Such a fit results in a $\chi^{2}$ = \LD{2527.72 for 2452} \LD{dof}. The resulting hydrogen column density is found to \LD{be $(11.2\pm0.2)\times10^{22}$ cm$^{-2}$, } much higher than that predicted from the dispersion measure (DM) of $\sim2\times10^{22}$ cm$^{-2}$ if we adopt a linear relationship of $N_{\rm{H}}$~(10$^{20}$~cm$^{-2}$)$=0.30_{-0.09}^{+0.13}$~DM~(pc~cm$^{-3}$) \citep{HeNK2013, KaruppusamyDK2020}. This is consistent with other sources near the Galactic center and suggests that a significant part of X-ray absorption could be contributed by molecular clouds rather than neutral hydrogen atoms \citep{BaumgartnerM2006, WillingaleSB2013}. \LD {We show the best fit models together with the first and the last X-ray spectra in Figure \ref{fig:spdata} and the best-fit parameters in Table \ref{tab:bbodyrad}.  We further show the evolution of the inferred spectral parameters in Figure \ref{fig:spdata}. The flux decayed to roughly half of the initial value. The temperature, however, does not show a clear variability and seems to agree within the statistical uncertainties of the individual measurements. We tested linking this parameter throughout all the observations. Such a fit yield an average kT=$1.05\pm0.01$ and a $\chi^{2}$ = 2619.63 for 2497 \LD{dof}. The apparent radius shows a decrease of $\sim60$\% regardless of whether the temperature is linked or not.}

The quiescent emission from \src\ is not detectable in archival X-ray observations. We used the deepest \emph{XMM-Newton} observation (ObsID: 0800910101) to estimate the bolometric 3$\sigma$ upper limit of the quiescent luminosity as $7\times10^{32}$--$1.7\times10^{34}$~\lumcgs. This range contains the uncertainty in the distance of $4.8$--$8.1$ kpc, and the possible blackbody temperature range of $kT=0.3$--$0.5$ keV. Using this value, we can derive a limit to the luminosity increase as a factor of $>$10.  Note that the spectral evolution shown in Figure \ref{fig:spdata} indicates that the surface temperature of the source did not change significantly.  The flux decay is dominated by the decrease in the apparent emitting radius, which decreases by about 30\%. This finding supports a scenario where the outburst decay is caused by a shrinking hotspot due to the untwisting of magnetic field loops \citep{Beloborodov2013}.

\section{Discussion}\label{sec:discussion}

In this paper, we report the analysis of the \swift\ BAT-detected magnetar-like burst\LD{, which led to the discovery} of \src, and our subsequent early \nicer\ monitoring campaign. \nicer's flexibility and ease of scheduling allowed for a high observing cadence on the source throughout the first $\sim$\LD{100}~d since the discovery, starting just a few hours after the BAT detection.

The $\sim$10~ms duration of the hard X-ray burst, while on the very short end of typical magnetar bursts, is not unprecedented (e.g., 4U~0142+61, \citealt{CollazziKH2015}; PSR~J1119$-$6127, \citealt{GogusLK2016}). Moreover, the burst thermal nature, the temperature we derive, $kT\approx8$~keV, and the area size, $R\approx2.5$~km, are within the range of the spectral characteristics of the majority of magnetar-like short bursts and follow the expectation of emission from a trapped fireball near the surface of a magnetar. Hence, the bursting behavior of \src\ places it well within the magnetar family. The \BD{21} short bursts detected with \nicer\ are commensurate in duration with the BAT-detected burst.

\subsection{Post-Outburst Timing Evolution}

The post-outburst timing behavior is largely erratic, consistent with the large torque variations observed from magnetars and high-$B$ RPPs during outburst epochs \citep{DibKG2009, ArchibaldKT2016, ArchibaldKS2017}. \LD{During the first two weeks of our monitoring campaign, the source showed a large spin-up glitch and a likely spin-down glitch}. No gradual recovery is observed as shown in regular RPPs \citep{EspinozaLS2011} and even radiatively-silent glitches in magnetars \citep{DibK2014}. The sizes of these two glitches are extremely large even compared with those in other magnetars \citep{DibK2014}. \LD{This implies a substantial change in the kinetic energy that could be released in the form of electromagnetic waves.} However, similar to 70\% of glitches in magnetars, we did not see any radiative change accompanying the spin-up glitch of \src\ except for possible clustering of short bursts \citep{JanssenS2006, DibK2014, KaspiAB2014}. The lack of radiative variability may imply a recovery that is dictated by processes internal to the NS. The more erratic changes observed during magnetar outbursts, including the one we observe for \src, may point to variations dominated by external processes, likely close to the light cylinder where particle outflow could exert large torques on the star. This requires a coupling between the inner-crust, where the glitch occurs, and the external dipolar field lines. This condition may be achieved in high-$B$ sources \citep{HardingCK1999, ThompsonDW2000}.
 


The spin-down glitches are rarely seen in magnetars and never observed in regular pulsars. The first confirmed anti-glitch was observed in 1E~2259+586 with a size of $\Delta\nu=-4.5\times10^{-8}$ Hz and $\Delta\dot{\nu}=-2.7\times10^{-14}$ s$^{-2}$, occurring at the onset of a radiative outburst \citep{ArchibaldKN2013}. Similar behavior has been observed in SGR~1900+14 during a burst active epoch although a gradual slow down remained a possible explanation due to a $\sim80$-day gap \citep{WoodsKP1999}. For \src, we do not detect any additional enhancement in the persistent emission coincident with the epoch of the anti-glitch. Recently, 1E~2259+586 has shown a radiatively-quiet anti-glitch with a sudden spin-down amplitude $\delta_{\rm nu}=-8.1\times10^{-8}$~Hz \citep{YounesRB2020}.  However, unlike 1E~2259+586,  the anti-glitch in \src\ was very close to the start of a major outburst. A radiative change may have occurred with this anti-glitch but was insufficient to present itself above the high persistent flux during the outburst. The mechanism for radiatively silent anti-glitches remains unclear. It is possibly originated from the magnetosphere, but the particle acceleration is too weak to trigger radiative events under extra loading of plasma \citep{HardingCK1999}. An alternative explanation is that the anti-glitch is caused by the coupling of a superfluid component with a rotation frequency lower than the rest of the NS.  It is difficult to interpret why the detached superfluid component spins down much faster than the rest of the NS before the anti-glitch. 



It remains possible that the entire temporal evolution is caused by a series of rapid and non-instantaneous torque variability similar to the post-outburst behavior in SGR 1806$-$20 and 1E~1048$-$5937 \citep{WoodsKF2007, DibKG2009, DibK2014}. Right after the giant flare on 2004, the $\dot{\nu}$ of SGR 1806$-$20 changed rapidly between $\sim-2\times10^{-12}$~s$^{-2}$ and $\sim-1.5\times10^{-11}$~s$^{-2}$. Similarly, after the onset of the outburst in 2009, the $\dot{\nu}$ of 1E~1048$-$5937 oscillated between $\sim-2\times10^{-13}$~s$^{-2}$ and $\sim-2.8\times10^{-12}$~s$^{-2}$ for $\sim500$ days. The post-outburst $\dot{\nu}$ of \src\ varied between $\sim-4\times10^{-12}$~s$^{-2}$ and $\sim-6\times10^{-11}$~s$^{-2}$, which has a similar relative amplitude compared to that of SGR1806$-$20 and 1E~1048$-$5937 with a much larger size. This suggests that \src\ is one of the noisiest sources among magnetars and high-$B$ RPPs. 

\subsection{Flux Decay and Spectral Evolution}
The post-outburst spectral evolution shows that although the observed trend in the \LD{unabsorbed} flux is not completely monotonic, it decreases by about 55\% in \LD{102} days, from \LD{3.79 to 1.27} $\times10^{-11}$~\fluxcgs. Using the spectral parameters we also calculated the 0.3$-$10.0~keV thermal luminosity of \src. We fit the decaying trend in the flux with both the plateau-decay model \citep[see equation 12 in][]{EnotoSK2017} and the double exponential model \LD{\citep[see equation 1 in][]{CotiZelatiRP2018}.}

\LD{Since \nicer\ is a non-imaging instrument, uncertainties in the estimated background spectrum may have significant effects, especially at low flux levels. We noticed a $\sim4$\% systematic uncertainty by analyzing several background fields observed in 2020. Additionally, \cite{EspositoRB2020} reported the detection of a dust scattering halo that contributes 2\% of the source flux. To take both effects into account, we introduce a 5\% systematic uncertainty in our fits. With these considerations, these models provide broadly acceptable fits to data. Note that simpler models, including an exponential or a power-law decay, result in substantially worse fits.} 

\LD{The plateau-decay model assumes that the luminosity first shows a plateau-like slow decay phase which is then followed by a power-law like decay after a certain timescale. It is characterized by the luminosity at the onset of the outburst, $L_0$, the timescale of the plateau, $\tau_0$  and the power-law index, $p$ \citep[see equation 12 in][]{EnotoSK2017}.} We obtained a $\chi^{2}$=166 with 43 dof for the plateau-decay model. The best fit parameters and their 1$\sigma$ uncertainties are L$_{0}$=$(1.92\pm0.06)\times10^{35}$~\lumcgs, $\tau_0$=23$_{-7}^{+11}$ days, and $p$=0.51$\pm$0.08. Before this study, only SGR~0501$+$4516, SGR~0418$+$5729, and Swift~J1822$-$1606 have detectable $\tau_0$ of 15.9, 42.9 and 11.2 days, respectively \citep{EnotoSK2017}. These $\tau_0$ measurements are similar to the value we get for \src. However, for all of these sources, the slopes of the decay are significantly larger than what we infer, indicating a much faster decline for the other sources compared to \src. Similarly, the decreasing trend in the luminosity can also be modeled by a double exponential decay function following equation 1 in \citet{CotiZelatiRP2018}.  We found the normalization constants of individual components best match the data for A=1.51$_{-0.08}^{+0.06}\times$10$^{35}$~\lumcgs\ and B=0.49$\pm0.09\times$10$^{35}$~\lumcgs. The e-folding times are $\tau_{1}$=157$\pm13$ days and $\tau_{2}$=9$_\pm2$ days. The fit results in a $\chi^2$=161 with 42 dof. This model shows that the luminosity decrease has two components: one showing a rapid decay and another long term decay trend. The best fit values for the e-folding times are in agreement with similar results from \citet{CotiZelatiRP2018}, especially the values found for SGR~0501+4516. 


\LD{Note that especially at the late stages of the decay, the persistent flux of the source shows significant fluctuations. The fact that at least in some of these observations we also detect short bursts may imply that low-level activity is quasi-continuous during the outburst decay, which could be affecting the apparent persistent flux level.}

\begin{figure*}
\begin{minipage}{0.45\linewidth}
\includegraphics[width=0.99\textwidth]{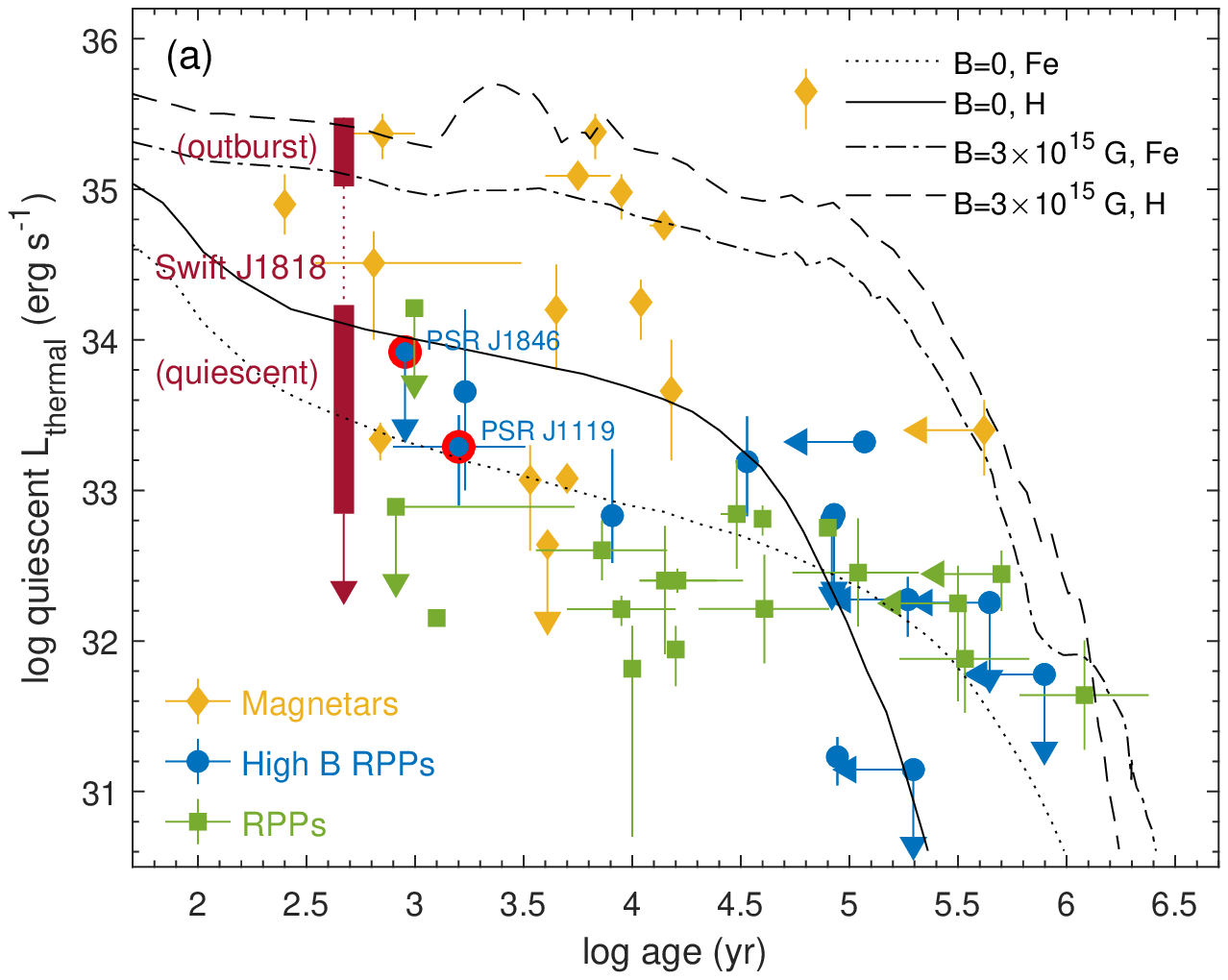} 
\end{minipage}
\hspace{0.0cm}
\begin{minipage}{0.45\linewidth}
\includegraphics[width=0.99\textwidth]{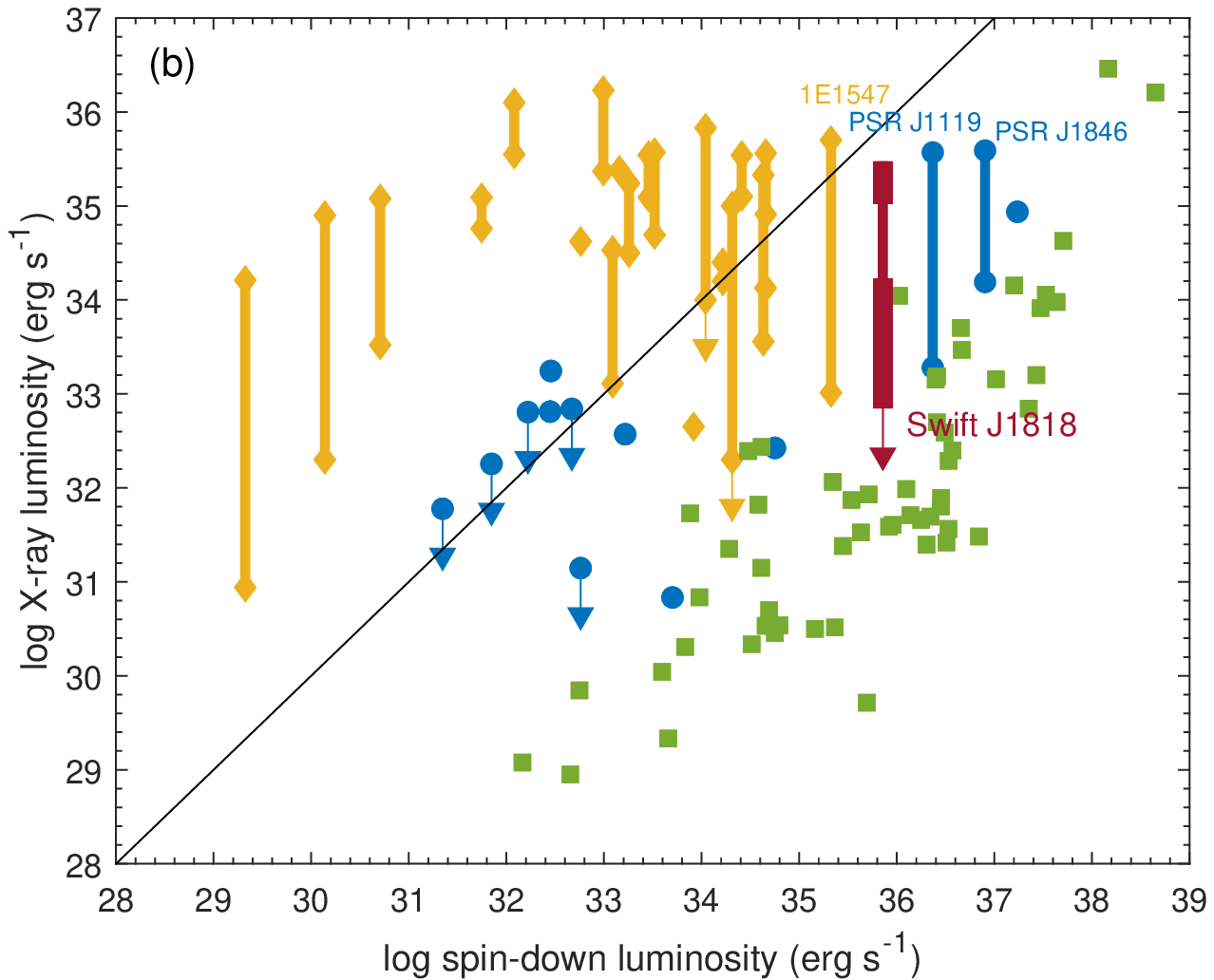} 
\end{minipage}
\caption{(a) The quiescent thermal luminosity observed in the soft X-ray band versus the age of magnetars, high-$B$ RPPs, RPPs, and \src. The data points are mainly taken from \citet{ShterninYH2011}, \citet{NgKH2012}, and \citet{ViganoRP2013} with a few updates listed in \citet{HuNT2017} and \citet{EnotoKS2019}. We use the age estimated from the supernova remnants if available, otherwise, we use the characteristic age. For theoretical cooling curves for a 1.4 $M_{\odot}$ NS with zero magnetic field and a Fe envelope, zero magnetic field and an H envelope, strong magnetic field and a Fe envelope, and strong magnetic field and an H envelope are adopted from \citet{ViganoRP2013}. The range of the upper limit for the quiescent luminosity of \src\ is labeled with a thick vertical bar, while the outburst luminosity is also plotted for reference. Two high $B$ RPPs that show magnetar-like behaviors are marked by red circles. (b) $L_{\rm{X}}$ versus $L_{\rm{sd}}$ for the aforementioned three types of pulsars. Black solid line denotes $L_{\rm{X}}=L_{\rm{sd}}$. The quiescent luminosity and outburst luminosity of transient magnetars and two high B-field RPPs are connected with solid lines. \label{fig:age_lumnosity}}
\end{figure*}

The broadband PF of the pulse profile of \src\ is $\gtrsim0.4$. Such a single-peaked pulse profile with a high PF may be difficult to \BD{produce with two} antipodal hotspots of equal brightness \citep{DedeoPN2001, HuNH2019}. Therefore, we suggest that the emission is dominated by a distorted hotspot on the surface of the NS. Emission from hot spots can be highly distorted, with anisotropy governed by the local $B$ field direction when the $B$ field is extremely high. The increase of the PF throughout our observing campaign reinforces the above idea that the outburst evolution is governed by the shrinking of a hotspot on the surface of the magnetar \citep{ReaIP2013, ZelatiRP2015, MongN2017}. Moreover, the hotspot could consist of two components with different temperatures. The boundary between the components could be much blurred instead of a sharp discontinuity \citep{DedeoPN2001}. These two components may have different shrinking timescales that result in the two timescales of the observed flux decay trend.   

A hard power-law spectral component above $\sim$10~keV has been reported from some of the persistently bright magnetars and from the early phases of transient outbursts, where the emission is thought to be radiated from the magnetosphere \citep[see, e.g.,][]{YounesKJ2017, ArchibaldSK2020}. Using the reported correlation between the soft and hard X-ray luminosity of known magnetars (equation 4 of \citealt{EnotoSK2017}) with the unabsorbed X-ray flux of \src, $\sim3\times 10^{-11}$~erg~s~cm$^{-2}$ observed in the soft band, we would expect the hard power-law flux at $\sim 2\times 10^{-11}$~erg~s~cm$^{-2}$ in the 15--60~keV band with a flat photon index of $\sim 0.8$ (equation 7 of \citealt{EnotoSK2017}). However, we did not find any evidence for such a hard power-law component in the soft \nicer\ spectrum. This is consistent with no detection above 15~keV with \emph{NuSTAR} and \emph{INTEGRAL} although a hard power-law component can be marginally seen below $\sim$20 keV \citep{BorgheseZR2020, EspositoRB2020}. This is in contrast to the prominent hard X-ray radiation in the 2009 outburst from a similar fast spinning and radio-emitting magnetar, 1E~1547.0$-$5408 \citep{EnotoNM2010}. 

\subsection{Nature of \src}
\src\ is a transient source showing timing properties between canonical magnetars and high-$B$ RPPs. Observations of low $B$-field magnetars and magnetar-like activity in high-$B$ RPPs hinted that magnetars could represent the high $B$ field tail of a single distribution \citep{KaspiM2005, Ho2013}. It has been suggested that magnetars have a high quiescent soft thermal luminosity that is powered by the dissipation of the strong magnetic field \citep{ThompsonD1995}. The magneto-thermal evolution model suggests that the key component is the toroidal magnetic field in the crust \citep{PonsMG2009, PernaP2011, ViganoRP2013}. This toroidal field cannot be inferred from the spin down. We plot the thermal luminosity of magnetars, high-$B$ RPPs, and several X-ray RPPs in Figure \ref{fig:age_lumnosity} (a). Several theoretical models with different magnetic field strengths and atmosphere composition are also plotted. Most magnetars have quiescent soft thermal luminosity above $\sim10^{34}$ \lumcgs\ except for transient radio magnetars. Canonical RPPs usually have luminosity lower than $10^{33}$~\lumcgs\ and several young high-$B$ RPPs are in between the RPPs and magnetars.

We overlay the estimated upper limit of the quiescent thermal luminosity of \src\ in Figure \ref{fig:age_lumnosity} (a). The luminosity of \src\ during the outburst, which is roughly the same as several bright persistent magnetars, e.g., 4U 0142+61 and 1RXS J170849.0$-$400910, is also plotted for reference. The upper limit of the quiescent luminosity of \src, although the uncertainty is large, occupies a similar region as the young high-$B$ RPPs J1846$-$0258 and J1119$-$6127, classifying \src\ in the same category. Moreover, the current estimate of $\nu$ and $\dot{\nu}$ may be affected by the glitch and the heavy timing noise similar to several young magnetars. The torque may be larger than the nominal value by $\gtrsim1$ order of magnitude at this early stage in the outburst. In several other magnetars, the torque decreases to the quiescent level on a timescale of a few months to ten years \citep{CamiloRH2016, YounesBK2017, ArchibaldSK2020}. Future long-term monitoring of the timing behavior of \src\ is necessary. If we adopted the $\dot{\nu}$ measurement from segment 3, the characteristic age of \src\ could be as high as $\gtrsim1000$ yr. This value is much older than that derived from the first two segments and comparable to that of PSR~J1846$-$0258. 

The detection of radio emission and the corresponding spectral index of \src\ provides another hint that \src\ can exhibit features of a regular RPP instead of a canonical magnetar \citep{KaruppusamyDK2020, EspositoRB2020, MajidPP2020}. Historically, magnetars are considered radio-silent NSs, where their $L_{\rm{X}}$ are higher than their $L_{\rm{sd}}$. The discovery of radio-emitting magnetars was a milestone that links the magnetars and RPPs. Their $L_{\rm{X}}$ drop to lower than their $L_{\rm{sd}}$ in quiescence. The high-$B$ RPPs J1846$-$0258 and J1119$-$6127 are two important samples to bridge radio-emitting magnetars and RPPs. They show magnetar bursts and X-ray outbursts, but the peak $L_{\rm{X}}$ remains lower than their $L_{\rm{sd}}$. We plotted the $L_{\rm{X}}$ versus the $L_{\rm{sd}}$ in Figure \ref{fig:age_lumnosity} (b) of all magnetars and RPPs together with the luminosity range of \src\ from expected quiescence to the outburst peak. The $L_{\rm{sd}}$ of \src\ is highest among the canonical magnetars and slightly lower than that of PSR~J1119$-$6127. Similar to high-$B$ RPPs, the peak $L_{\rm{X}}$ of \src\ remains lower than its $L_{\rm{sd}}$.  Moreover, radio-emitting magnetars show intermittent radio emission. On the contrary, the radio emission of PSR~J1119$-$6127 shut off during the early stages of its outburst onset \citep{MajidPD2017}. \src\ shows signs of both magnetar and radio pulsar populations and provides a crucial link between the two populations. Continued radio and X-ray monitoring of \src\ is critical to better understand the nature of this source.


\section{Summary}\label{sec:summary}
In this paper, we report the hard X-ray properties of the \src\ burst seen by \swift\ BAT, and the soft X-ray temporal/spectral evolution with \nicer. The profile and the spectral properties of the hard X-ray burst are in line with those from other magnetars. The subsequent \nicer\ monitoring suggests a long-term spin-down rate of $\dot{\nu}=-2.74\times10^{-11}$~s$^{-2}$ that implies an equatorial $B=2.7\times10^{14}$~G. The $L_{\rm{sd}}=7.9\times10^{35}$~\lumcgs\ is between that of typical magnetars and high-$B$ RPPs.  Moreover, we observed a glitch and a candidate anti-glitch during the \nicer\ monitoring. These two glitches have the largest size among glitches in magnetars but we do not observe significant radiative events associated with them. From spectral analysis, we suggest that persistent X-rays from \src\ are dominated by thermal emission of a hotspot on the surface. The increase of the PF and the two-stage flux decay can be interpreted as the shrinking of the hotspot size, which has two components with different shrinking timescales. Finally, we suggest that \src\ is an important link that bridges magnetars and high-$B$ RPPs, based on its timing properties and low X-ray luminosity.

\section*{ACKNOWLEDGMENTS}
We thank Professor Victoria Kaspi for useful discussions and the anonymous reviewer for valuable comments that improved this paper. This work was supported by the National Aeronautics and Space Administration (NASA) through the \nicer\ mission and the Astrophysics Explorers Program. The \nicer\ observation campaign was performed under the NICER GO2 program 3056 “Magnetic Energy Dissipation of Magnetar Outbursts Studied via Multiwavelength Follow-up Observation” (PI: Teruaki Enoto). This work partly made use of data supplied by the UK \swift\ Science Data Centre at the University of Leicester, and observations obtained with \emph{XMM-Newton} and the ESA science mission with instruments and contributions directly funded by the ESA member states and NASA. C.-P.H. acknowledges support from the Japan Society for the Promotion of Science (JSPS; ID: P18318). T.G. has been supported in part by the Royal Society Newton Advanced Fellowship, NAF$\backslash$R2$\backslash$180592, and the Turkish Republic, Directorate of Presidential Strategy and Budget project, 2016K121370. T.E. has been supported by the JSPS/MEXT KAKENHI grant numbers 16H02198 18H01246 and the Hakubi projects of Kyoto University and RIKEN.  C.M. is supported by the NASA Postdoctoral Program at the Marshall Space Flight Center, administered by Universities Space Research Association under contract with NASA. Z.W. acknowledges support from the NASA postdoctoral program. A portion of this research was carried out at the Jet Propulsion Laboratory, California Institute of Technology, under a contract with NASA. This work has made use of the NASA Astrophysics Data System.

\facilities{NICER}
\software{HEASoft, XSPEC}

\bibliographystyle{aasjournal}

\begin{thebibliography}{}
\expandafter\ifx\csname natexlab\endcsname\relax\def\natexlab#1{#1}\fi
\providecommand{\url}[1]{\href{#1}{#1}}
\providecommand{\dodoi}[1]{doi:~\href{http://doi.org/#1}{\nolinkurl{#1}}}
\providecommand{\doeprint}[1]{\href{http://ascl.net/#1}{\nolinkurl{http://ascl.net/#1}}}
\providecommand{\doarXiv}[1]{\href{https://arxiv.org/abs/#1}{\nolinkurl{https://arxiv.org/abs/#1}}}

\bibitem[{{An} {et~al.}(2015){An}, {Archibald}, {Hasco{\"e}t}, {Kaspi},
  {Beloborodov}, {Archibald}, {Beardmore}, {Boggs}, {Christensen}, {Craig},
  {Gehrels}, {Hailey}, {Harrison}, {Kennea}, {Kouveliotou}, {Stern}, {Younes},
  \& {Zhang}}]{AnAH2015}
{An}, H., {Archibald}, R.~F., {Hasco{\"e}t}, R., {et~al.} 2015, \apj, 807, 93,
  \dodoi{10.1088/0004-637X/807/1/93}

\bibitem[{{Archibald} {et~al.}(2017){Archibald}, {Kaspi}, {Scholz},
  {Beardmore}, {Gehrels}, \& {Kennea}}]{ArchibaldKS2017}
{Archibald}, R.~F., {Kaspi}, V.~M., {Scholz}, P., {et~al.} 2017, \apj, 834,
  163, \dodoi{10.3847/1538-4357/834/2/163}

\bibitem[{{Archibald} {et~al.}(2016){Archibald}, {Kaspi}, {Tendulkar}, \&
  {Scholz}}]{ArchibaldKT2016}
{Archibald}, R.~F., {Kaspi}, V.~M., {Tendulkar}, S.~P., \& {Scholz}, P. 2016,
  \apjl, 829, L21, \dodoi{10.3847/2041-8205/829/1/L21}

\bibitem[{{Archibald} {et~al.}(2020){Archibald}, {Scholz}, {Kaspi},
  {Tendulkar}, \& {Beardmore}}]{ArchibaldSK2020}
{Archibald}, R.~F., {Scholz}, P., {Kaspi}, V.~M., {Tendulkar}, S.~P., \&
  {Beardmore}, A.~P. 2020, \apj, 889, 160, \dodoi{10.3847/1538-4357/ab660c}

\bibitem[{{Archibald} {et~al.}(2013){Archibald}, {Kaspi}, {Ng},
  {Gourgouliatos}, {Tsang}, {Scholz}, {Beardmore}, {Gehrels}, \&
  {Kennea}}]{ArchibaldKN2013}
{Archibald}, R.~F., {Kaspi}, V.~M., {Ng}, C.-Y., {et~al.} 2013, \nat, 497, 591,
  \dodoi{10.1038/nature12159}

\bibitem[{{Arnaud}(1996)}]{Arnaud1996}
{Arnaud}, K.~A. 1996, in Astronomical Society of the Pacific Conference Series,
  Vol. 101, Astronomical Data Analysis Software and Systems V, ed. G.~H.
  {Jacoby} \& J.~{Barnes}, 17

\bibitem[{{Barthelmy} {et~al.}(2005){Barthelmy}, {Barbier}, {Cummings},
  {Fenimore}, {Gehrels}, {Hullinger}, {Krimm}, {Markwardt}, {Palmer},
  {Parsons}, {Sato}, {Suzuki}, {Takahashi}, {Tashiro}, \&
  {Tueller}}]{BarthelmyBC2005}
{Barthelmy}, S.~D., {Barbier}, L.~M., {Cummings}, J.~R., {et~al.} 2005, \ssr,
  120, 143, \dodoi{10.1007/s11214-005-5096-3}

\bibitem[{{Baumgartner} \& {Mushotzky}(2006)}]{BaumgartnerM2006}
{Baumgartner}, W.~H., \& {Mushotzky}, R.~F. 2006, \apj, 639, 929,
  \dodoi{10.1086/499619}

\bibitem[{{Beloborodov}(2013)}]{Beloborodov2013}
{Beloborodov}, A.~M. 2013, \apj, 777, 114, \dodoi{10.1088/0004-637X/777/2/114}

\bibitem[{{Borghese} {et~al.}(2020){Borghese}, {Zelati}, {Rea}, {Israel},
  {Esposito}, {Gotz}, {Savchenko}, {Tiengo}, {Mereghetti}, {Pintore},
  {Ridolfi}, {Rodriguez}, \& {Vigano}}]{BorgheseZR2020}
{Borghese}, A., {Zelati}, F.~C., {Rea}, N., {et~al.} 2020, The Astronomer's
  Telegram, 13569, 1

\bibitem[{{Camilo} {et~al.}(2007){Camilo}, {Ransom}, {Halpern}, \&
  {Reynolds}}]{CamiloRH2007}
{Camilo}, F., {Ransom}, S.~M., {Halpern}, J.~P., \& {Reynolds}, J. 2007, \apjl,
  666, L93, \dodoi{10.1086/521826}

\bibitem[{{Camilo} {et~al.}(2016){Camilo}, {Ransom}, {Halpern}, {Alford},
  {Cognard}, {Reynolds}, {Johnston}, {Sarkissian}, \& {van
  Straten}}]{CamiloRH2016}
{Camilo}, F., {Ransom}, S.~M., {Halpern}, J.~P., {et~al.} 2016, \apj, 820, 110,
  \dodoi{10.3847/0004-637X/820/2/110}

\bibitem[{{Champion} {et~al.}(2020){Champion}, {Desvignes}, {Jankowski},
  {Karuppusamy}, {Keith}, {Kouveliotou}, {Kramer}, {Lyne}, {Mickaliger},
  {O'Connor}, {Porayko}, {Rajwade}, {Stappers}, {Torne}, {van der Horst}, \&
  {Weltevrede}}]{ChampionDJ2020}
{Champion}, D., {Desvignes}, G., {Jankowski}, F., {et~al.} 2020, The
  Astronomer's Telegram, 13559, 1

\bibitem[{Cleveland \& Loader(1996)}]{ClevelandL1996}
Cleveland, W.~S., \& Loader, C. 1996, in Statistical Theory and Computational
  Aspects of Smoothing, ed. W.~H{\"a}rdle \& M.~G. Schimek (Heidelberg:
  Physica-Verlag HD), 10--49

\bibitem[{{Collazzi} {et~al.}(2015){Collazzi}, {Kouveliotou}, {van der Horst},
  {Younes}, {Kaneko}, {G{\"o}{\u{g}}{\"u}{\textcommabelow s}}, {Lin}, {Granot},
  {Finger}, {Chaplin}, {Huppenkothen}, {Watts}, {von Kienlin}, {Baring},
  {Gruber}, {Bhat}, {Gibby}, {Gehrels}, {McEnery}, {van der Klis}, \&
  {Wijers}}]{CollazziKH2015}
{Collazzi}, A.~C., {Kouveliotou}, C., {van der Horst}, A.~J., {et~al.} 2015,
  \apjs, 218, 11, \dodoi{10.1088/0067-0049/218/1/11}

\bibitem[{{Coti Zelati} {et~al.}(2018){Coti Zelati}, {Rea}, {Pons}, {Campana},
  \& {Esposito}}]{CotiZelatiRP2018}
{Coti Zelati}, F., {Rea}, N., {Pons}, J.~A., {Campana}, S., \& {Esposito}, P.
  2018, \mnras, 474, 961, \dodoi{10.1093/mnras/stx2679}

\bibitem[{{Coti Zelati} {et~al.}(2015){Coti Zelati}, {Rea}, {Papitto},
  {Vigan{\`o}}, {Pons}, {Turolla}, {Esposito}, {Haggard}, {Baganoff}, {Ponti},
  {Israel}, {Campana}, {Torres}, {Tiengo}, {Mereghetti}, {Perna}, {Zane},
  {Mignani}, {Possenti}, \& {Stella}}]{ZelatiRP2015}
{Coti Zelati}, F., {Rea}, N., {Papitto}, A., {et~al.} 2015, \mnras, 449, 2685,
  \dodoi{10.1093/mnras/stv480}

\bibitem[{{DeDeo} {et~al.}(2001){DeDeo}, {Psaltis}, \& {Narayan}}]{DedeoPN2001}
{DeDeo}, S., {Psaltis}, D., \& {Narayan}, R. 2001, \apj, 559, 346,
  \dodoi{10.1086/322283}

\bibitem[{{Dib} \& {Kaspi}(2014)}]{DibK2014}
{Dib}, R., \& {Kaspi}, V.~M. 2014, \apj, 784, 37,
  \dodoi{10.1088/0004-637X/784/1/37}

\bibitem[{{Dib} {et~al.}(2009){Dib}, {Kaspi}, \& {Gavriil}}]{DibKG2009}
{Dib}, R., {Kaspi}, V.~M., \& {Gavriil}, F.~P. 2009, \apj, 702, 614,
  \dodoi{10.1088/0004-637X/702/1/614}

\bibitem[{{Duncan} \& {Thompson}(1992)}]{DuncanT1992}
{Duncan}, R.~C., \& {Thompson}, C. 1992, \apjl, 392, L9, \dodoi{10.1086/186413}

\bibitem[{{Enoto} {et~al.}(2019){Enoto}, {Kisaka}, \& {Shibata}}]{EnotoKS2019}
{Enoto}, T., {Kisaka}, S., \& {Shibata}, S. 2019, Reports on Progress in
  Physics, 82, 106901, \dodoi{10.1088/1361-6633/ab3def}

\bibitem[{{Enoto} {et~al.}(2010){Enoto}, {Nakazawa}, {Makishima}, {Nakagawa},
  {Sakamoto}, {Ohno}, {Takahashi}, {Terada}, {Yamaoka}, {Murakami}, \&
  {Takahashi}}]{EnotoNM2010}
{Enoto}, T., {Nakazawa}, K., {Makishima}, K., {et~al.} 2010, \pasj, 62, 475,
  \dodoi{10.1093/pasj/62.2.475}

\bibitem[{Enoto {et~al.}(2017)Enoto, Shibata, Kitaguchi, Suwa, Uchide,
  Nishioka, Kisaka, Nakano, Murakami, \& Makishima}]{EnotoSK2017}
Enoto, T., Shibata, S., Kitaguchi, T., {et~al.} 2017, The Astrophysical Journal
  Supplement Series, 231, 8, \dodoi{10.3847/1538-4365/aa6f0a}

\bibitem[{{Enoto} {et~al.}(2020){Enoto}, {Sakamoto}, {Younes}, {Hu}, {Ho},
  {Gendreau}, {Arzoumanian}, {Guver}, {Guillot}, {Altamirano}, {Ray}, {Ng},
  {Chakrabarty}, {Jaisawal}, \& {Bogdanov}}]{EnotoSY2020}
{Enoto}, T., {Sakamoto}, T., {Younes}, G., {et~al.} 2020, The Astronomer's
  Telegram, 13551, 1

\bibitem[{{Espinoza} {et~al.}(2011){Espinoza}, {Lyne}, {Stappers}, \&
  {Kramer}}]{EspinozaLS2011}
{Espinoza}, C.~M., {Lyne}, A.~G., {Stappers}, B.~W., \& {Kramer}, M. 2011,
  \mnras, 414, 1679, \dodoi{10.1111/j.1365-2966.2011.18503.x}

\bibitem[{{Esposito} {et~al.}(2020){Esposito}, {Rea}, {Borghese}, {Coti
  Zelati}, {Vigan{\`o}}, {Israel}, {Tiengo}, {Ridolfi}, {Possenti}, {Burgay},
  {G{\"o}tz}, {Pintore}, {Stella}, {Dehman}, {Ronchi}, {Campana},
  {Garcia-Garcia}, {Graber}, {Mereghetti}, {Perna}, {Turolla}, \&
  {Zane}}]{EspositoRB2020}
{Esposito}, P., {Rea}, N., {Borghese}, A., {et~al.} 2020, arXiv e-prints,
  arXiv:2004.04083.
\newblock \doarXiv{2004.04083}

\bibitem[{{Evans} {et~al.}(2020){Evans}, {Gropp}, {Kennea}, {Klingler}, {Laha},
  {Lien}, {Page}, {Sakamoto}, {Tohuvavohu}, \& {Neil Gehrels Swift Observatory
  Team}}]{EvansGK2020}
{Evans}, P.~A., {Gropp}, J.~D., {Kennea}, J.~A., {et~al.} 2020, GRB Coordinates
  Network, 27373, 1

\bibitem[{{Gavriil} {et~al.}(2008){Gavriil}, {Gonzalez}, {Gotthelf}, {Kaspi},
  {Livingstone}, \& {Woods}}]{GavriilGG2008}
{Gavriil}, F.~P., {Gonzalez}, M.~E., {Gotthelf}, E.~V., {et~al.} 2008, Science,
  319, 1802, \dodoi{10.1126/science.1153465}

\bibitem[{{Gehrels} {et~al.}(2004){Gehrels}, {Chincarini}, {Giommi}, {Mason},
  {Nousek}, {Wells}, {White}, {Barthelmy}, {Burrows}, {Cominsky}, {Hurley},
  {Marshall}, {M{\'e}sz{\'a}ros}, {Roming}, {Angelini}, {Barbier}, {Belloni},
  {Campana}, {Caraveo}, {Chester}, {Citterio}, {Cline}, {Cropper}, {Cummings},
  {Dean}, {Feigelson}, {Fenimore}, {Frail}, {Fruchter}, {Garmire}, {Gendreau},
  {Ghisellini}, {Greiner}, {Hill}, {Hunsberger}, {Krimm}, {Kulkarni}, {Kumar},
  {Lebrun}, {Lloyd-Ronning}, {Markwardt}, {Mattson}, {Mushotzky}, {Norris},
  {Osborne}, {Paczynski}, {Palmer}, {Park}, {Parsons}, {Paul}, {Rees},
  {Reynolds}, {Rhoads}, {Sasseen}, {Schaefer}, {Short}, {Smale}, {Smith},
  {Stella}, {Tagliaferri}, {Takahashi}, {Tashiro}, {Townsley}, {Tueller},
  {Turner}, {Vietri}, {Voges}, {Ward}, {Willingale}, {Zerbi}, \&
  {Zhang}}]{GehrelsCG2004}
{Gehrels}, N., {Chincarini}, G., {Giommi}, P., {et~al.} 2004, \apj, 611, 1005,
  \dodoi{10.1086/422091}

\bibitem[{{G{\"o}{\u g}{\"u}{\c s}} {et~al.}(2016){G{\"o}{\u g}{\"u}{\c s}},
  {Lin}, {Kaneko}, {Kouveliotou}, {Watts}, {Chakraborty}, {Alpar},
  {Huppenkothen}, {Roberts}, {Younes}, \& {van der Horst}}]{GogusLK2016}
{G{\"o}{\u g}{\"u}{\c s}}, E., {Lin}, L., {Kaneko}, Y., {et~al.} 2016, \apjl,
  829, L25, \dodoi{10.3847/2041-8205/829/2/L25}

\bibitem[{{Harding} {et~al.}(1999){Harding}, {Contopoulos}, \&
  {Kazanas}}]{HardingCK1999}
{Harding}, A.~K., {Contopoulos}, I., \& {Kazanas}, D. 1999, \apjl, 525, L125,
  \dodoi{10.1086/312339}

\bibitem[{{He} {et~al.}(2013){He}, {Ng}, \& {Kaspi}}]{HeNK2013}
{He}, C., {Ng}, C.-Y., \& {Kaspi}, V.~M. 2013, \apj, 768, 64,
  \dodoi{10.1088/0004-637X/768/1/64}

\bibitem[{HEASARC(2014)}]{Heasarc2014}
HEASARC. 2014, {HEAsoft: Unified Release of FTOOLS and XANADU}.
\newblock \doeprint{1408.004}

\bibitem[{{Ho}(2013)}]{Ho2013}
{Ho}, W.~C.~G. 2013, \mnras, 429, 113, \dodoi{10.1093/mnras/sts317}

\bibitem[{{Hu} {et~al.}(2019){Hu}, {Ng}, \& {Ho}}]{HuNH2019}
{Hu}, C.-P., {Ng}, C.-Y., \& {Ho}, W.~C.~G. 2019, \mnras, 485, 4274,
  \dodoi{10.1093/mnras/stz513}

\bibitem[{{Hu} {et~al.}(2017){Hu}, {Ng}, {Takata}, {Shannon}, \&
  {Johnston}}]{HuNT2017}
{Hu}, C.-P., {Ng}, C.-Y., {Takata}, J., {Shannon}, R.~M., \& {Johnston}, S.
  2017, \apj, 838, 156

\bibitem[{{Janssen} \& {Stappers}(2006)}]{JanssenS2006}
{Janssen}, G.~H., \& {Stappers}, B.~W. 2006, \aap, 457, 611,
  \dodoi{10.1051/0004-6361:20065267}

\bibitem[{{Karuppusamy} {et~al.}(2020){Karuppusamy}, {Desvignes}, {Kramer},
  {Porayko}, {Champion}, {Torne}, {Stappers}, {van der Horst}, {Kouveliotou},
  \& {O'Connor}}]{KaruppusamyDK2020}
{Karuppusamy}, R., {Desvignes}, G., {Kramer}, M., {et~al.} 2020, The
  Astronomer's Telegram, 13553, 1

\bibitem[{{Kaspi} \& {Beloborodov}(2017)}]{KaspiB2017}
{Kaspi}, V.~M., \& {Beloborodov}, A.~M. 2017, \araa, 55, 261,
  \dodoi{10.1146/annurev-astro-081915-023329}

\bibitem[{{Kaspi} \& {McLaughlin}(2005)}]{KaspiM2005}
{Kaspi}, V.~M., \& {McLaughlin}, M.~A. 2005, \apjl, 618, L41,
  \dodoi{10.1086/427628}

\bibitem[{{Kaspi} {et~al.}(2014){Kaspi}, {Archibald}, {Bhalerao}, {Dufour},
  {Gotthelf}, {An}, {Bachetti}, {Beloborodov}, {Boggs}, {Christensen}, {Craig},
  {Grefenstette}, {Hailey}, {Harrison}, {Kennea}, {Kouveliotou}, {Madsen},
  {Mori}, {Markwardt}, {Stern}, {Vogel}, \& {Zhang}}]{KaspiAB2014}
{Kaspi}, V.~M., {Archibald}, R.~F., {Bhalerao}, V., {et~al.} 2014, \apj, 786,
  84, \dodoi{10.1088/0004-637X/786/2/84}

\bibitem[{{Levin} {et~al.}(2010){Levin}, {Bailes}, {Bates}, {Bhat}, {Burgay},
  {Burke-Spolaor}, {D'Amico}, {Johnston}, {Keith}, {Kramer}, {Milia},
  {Possenti}, {Rea}, {Stappers}, \& {van Straten}}]{LevinBB2010}
{Levin}, L., {Bailes}, M., {Bates}, S., {et~al.} 2010, \apjl, 721, L33,
  \dodoi{10.1088/2041-8205/721/1/L33}

\bibitem[{{Livingstone} {et~al.}(2009){Livingstone}, {Ransom}, {Camilo},
  {Kaspi}, {Lyne}, {Kramer}, \& {Stairs}}]{LivingstoneRC2009}
{Livingstone}, M.~A., {Ransom}, S.~M., {Camilo}, F., {et~al.} 2009, \apj, 706,
  1163, \dodoi{10.1088/0004-637X/706/2/1163}

\bibitem[{{Majid} {et~al.}(2017){Majid}, {Pearlman}, {Dobreva}, {Horiuchi},
  {Kocz}, {Lippuner}, \& {Prince}}]{MajidPD2017}
{Majid}, W.~A., {Pearlman}, A.~B., {Dobreva}, T., {et~al.} 2017, \apjl, 834,
  L2, \dodoi{10.3847/2041-8213/834/1/L2}

\bibitem[{{Majid} {et~al.}(2020){Majid}, {Pearlman}, {Prince}, {Naudet},
  {Kocz}, {Horiuchi}, {Enoto}, \& {Younes}}]{MajidPP2020}
{Majid}, W.~A., {Pearlman}, A.~B., {Prince}, T.~A., {et~al.} 2020, The
  Astronomer's Telegram, 13649, 1

\bibitem[{{Malacaria} {et~al.}(2020){Malacaria}, {Roberts}, {Veres}, \&
  {Wilson-Hodge}}]{MalacariaRV2020}
{Malacaria}, C., {Roberts}, O.~J., {Veres}, P., \& {Wilson-Hodge}, C. 2020, The
  Astronomer's Telegram, 13555, 1

\bibitem[{{Mong} \& {Ng}(2018)}]{MongN2017}
{Mong}, Y.-L., \& {Ng}, C.-Y. 2018, \apj, 852, 86,
  \dodoi{10.3847/1538-4357/aa9e90}

\bibitem[{{Ng} \& {Kaspi}(2011)}]{NgK2011}
{Ng}, C.-Y., \& {Kaspi}, V.~M. 2011, in AIP Conf. Ser. 1379, Astrophysics of
  Neutron Stars 2010: A Conference in Honor of M. Ali Alpar, ed. E.~{G{\"o}{\u
  g}{\"u}{\c s}}, T.~{Belloni}, \& {\"U}.~{Ertan}~(Melville, NY:~AIP), 60,
  \dodoi{10.1063/1.3629486}

\bibitem[{{Ng} {et~al.}(2012){Ng}, {Kaspi}, {Ho}, {Weltevrede}, {Bogdanov},
  {Shannon}, \& {Gonzalez}}]{NgKH2012}
{Ng}, C.-Y., {Kaspi}, V.~M., {Ho}, W.~C.~G., {et~al.} 2012, \apj, 761, 65,
  \dodoi{10.1088/0004-637X/761/1/65}

\bibitem[{{Paczynski}(1992)}]{Paczynski1992}
{Paczynski}, B. 1992, \actaa, 42, 145

\bibitem[{{Palmer}(2020)}]{Palmer2020}
{Palmer}, D.~M. 2020, The Astronomer's Telegram, 13675, 1

\bibitem[{{Perna} \& {Pons}(2011)}]{PernaP2011}
{Perna}, R., \& {Pons}, J.~A. 2011, \apjl, 727, L51,
  \dodoi{10.1088/2041-8205/727/2/L51}

\bibitem[{{Pons} {et~al.}(2009){Pons}, {Miralles}, \& {Geppert}}]{PonsMG2009}
{Pons}, J.~A., {Miralles}, J.~A., \& {Geppert}, U. 2009, \aap, 496, 207,
  \dodoi{10.1051/0004-6361:200811229}

\bibitem[{{Rea} {et~al.}(2016){Rea}, {Borghese}, {Esposito}, {Coti Zelati},
  {Bachetti}, {Israel}, \& {De Luca}}]{ReaBE2016}
{Rea}, N., {Borghese}, A., {Esposito}, P., {et~al.} 2016, \apjl, 828, L13,
  \dodoi{10.3847/2041-8205/828/1/L13}

\bibitem[{{Rea} {et~al.}(2010){Rea}, {Esposito}, {Turolla}, {Israel}, {Zane},
  {Stella}, {Mereghetti}, {Tiengo}, {G{\"o}tz}, {G{\"o}{\u g}{\"u}{\c s}}, \&
  {Kouveliotou}}]{ReaET2010}
{Rea}, N., {Esposito}, P., {Turolla}, R., {et~al.} 2010, Science, 330, 944,
  \dodoi{10.1126/science.1196088}

\bibitem[{{Rea} {et~al.}(2013){Rea}, {Israel}, {Pons}, {Turolla}, {Vigan{\`o}},
  {Zane}, {Esposito}, {Perna}, {Papitto}, {Terreran}, {Tiengo}, {Salvetti},
  {Girart}, {Palau}, {Possenti}, {Burgay}, {G{\"o}{\u g}{\"u}{\c s}},
  {Caliandro}, {Kouveliotou}, {G{\"o}tz}, {Mignani}, {Ratti}, \&
  {Stella}}]{ReaIP2013}
{Rea}, N., {Israel}, G.~L., {Pons}, J.~A., {et~al.} 2013, \apj, 770, 65,
  \dodoi{10.1088/0004-637X/770/1/65}

\bibitem[{{Scholz} \& {Chime/Frb Collaboration}(2020)}]{ScholzC2020}
{Scholz}, P., \& {Chime/Frb Collaboration}. 2020, The Astronomer's Telegram,
  13681, 1

\bibitem[{{Shannon} \& {Johnston}(2013)}]{ShannonJ2013}
{Shannon}, R.~M., \& {Johnston}, S. 2013, \mnras, 435, L29,
  \dodoi{10.1093/mnrasl/slt088}

\bibitem[{{Shternin} {et~al.}(2011){Shternin}, {Yakovlev}, {Heinke}, {Ho}, \&
  {Patnaude}}]{ShterninYH2011}
{Shternin}, P.~S., {Yakovlev}, D.~G., {Heinke}, C.~O., {Ho}, W.~C.~G., \&
  {Patnaude}, D.~J. 2011, \mnras, 412, L108,
  \dodoi{10.1111/j.1745-3933.2011.01015.x}

\bibitem[{{Thompson} \& {Duncan}(1995)}]{ThompsonD1995}
{Thompson}, C., \& {Duncan}, R.~C. 1995, \mnras, 275, 255,
  \dodoi{10.1093/mnras/275.2.255}

\bibitem[{{Thompson} {et~al.}(2000){Thompson}, {Duncan}, {Woods},
  {Kouveliotou}, {Finger}, \& {van Paradijs}}]{ThompsonDW2000}
{Thompson}, C., {Duncan}, R.~C., {Woods}, P.~M., {et~al.} 2000, \apj, 543, 340,
  \dodoi{10.1086/317072}

\bibitem[{{Vigan{\`o}} {et~al.}(2013){Vigan{\`o}}, {Rea}, {Pons}, {Perna},
  {Aguilera}, \& {Miralles}}]{ViganoRP2013}
{Vigan{\`o}}, D., {Rea}, N., {Pons}, J.~A., {et~al.} 2013, \mnras, 434, 123,
  \dodoi{10.1093/mnras/stt1008}

\bibitem[{{Willingale} {et~al.}(2013){Willingale}, {Starling}, {Beardmore},
  {Tanvir}, \& {O'Brien}}]{WillingaleSB2013}
{Willingale}, R., {Starling}, R.~L.~C., {Beardmore}, A.~P., {Tanvir}, N.~R., \&
  {O'Brien}, P.~T. 2013, \mnras, 431, 394, \dodoi{10.1093/mnras/stt175}

\bibitem[{{Wilms} {et~al.}(2000){Wilms}, {Allen}, \& {McCray}}]{WilmsAM2000}
{Wilms}, J., {Allen}, A., \& {McCray}, R. 2000, \apj, 542, 914,
  \dodoi{10.1086/317016}

\bibitem[{{Woods} {et~al.}(2007){Woods}, {Kouveliotou}, {Finger}, {G{\"o}{\v
  g}{\"u}{\c s}}, {Wilson}, {Patel}, {Hurley}, \& {Swank}}]{WoodsKF2007}
{Woods}, P.~M., {Kouveliotou}, C., {Finger}, M.~H., {et~al.} 2007, \apj, 654,
  470, \dodoi{10.1086/507459}

\bibitem[{{Woods} {et~al.}(1999){Woods}, {Kouveliotou}, {van Paradijs},
  {Hurley}, {Kippen}, {Finger}, {Briggs}, {Dieters}, \&
  {Fishman}}]{WoodsKP1999}
{Woods}, P.~M., {Kouveliotou}, C., {van Paradijs}, J., {et~al.} 1999, \apjl,
  519, L139, \dodoi{10.1086/312124}

\bibitem[{{Younes} {et~al.}(2017{\natexlab{a}}){Younes}, {Baring},
  {Kouveliotou}, {Harding}, {Donovan}, {G{\"o}{\u{g}}{\"u}{\textcommabelow s}},
  {Kaspi}, \& {Granot}}]{YounesBK2017}
{Younes}, G., {Baring}, M.~G., {Kouveliotou}, C., {et~al.} 2017{\natexlab{a}},
  \apj, 851, 17, \dodoi{10.3847/1538-4357/aa96fd}

\bibitem[{{Younes} {et~al.}(2020{\natexlab{a}}){Younes}, {Ray}, {Baring},
  {Kouveliotou}, {Fletcher}, {Wadiasingh}, {Harding}, \&
  {Goldstein}}]{YounesRB2020}
{Younes}, G., {Ray}, P.~S., {Baring}, M.~G., {et~al.} 2020{\natexlab{a}},
  \apjl, 896, L42, \dodoi{10.3847/2041-8213/ab9a48}

\bibitem[{{Younes} {et~al.}(2016){Younes}, {Kouveliotou}, {Kargaltsev}, {Gill},
  {Granot}, {Watts}, {Gelfand}, {Baring}, {Harding}, {Pavlov}, {van der Horst},
  {Huppenkothen}, {G{\"o}{\u g}{\"u}{\c s}}, {Lin}, \&
  {Roberts}}]{YounesKK2016}
{Younes}, G., {Kouveliotou}, C., {Kargaltsev}, O., {et~al.} 2016, \apj, 824,
  138, \dodoi{10.3847/0004-637X/824/2/138}

\bibitem[{{Younes} {et~al.}(2017{\natexlab{b}}){Younes}, {Kouveliotou},
  {Jaodand}, {Baring}, {van der Horst}, {Harding}, {Hessels}, {Gehrels},
  {Gill}, {Huppenkothen}, {Granot}, {G{\"o}{\u{g}}{\"u}{\textcommabelow s}}, \&
  {Lin}}]{YounesKJ2017}
{Younes}, G., {Kouveliotou}, C., {Jaodand}, A., {et~al.} 2017{\natexlab{b}},
  \apj, 847, 85, \dodoi{10.3847/1538-4357/aa899a}

\bibitem[{{Younes} {et~al.}(2020{\natexlab{b}}){Younes}, {Guver}, {Enoto},
  {Arzoumanian}, {Gendreau}, {Hu}, {Ray}, {Kouveliotou}, {Guillot}, {Ho},
  {Ferrara}, \& {Malacaria}}]{YounesGE2020}
{Younes}, G., {Guver}, T., {Enoto}, T., {et~al.} 2020{\natexlab{b}}, The
  Astronomer's Telegram, 13678, 1

\bibitem[{{Zhang} {et~al.}(2020){Zhang}, {Tuo}, {Xiong}, {Li}, {Xiao}, {Jia},
  {Li}, {Ge}, {Luo}, {Li}, {Cai}, {Tan}, {Xue}, {Lu}, {Song}, {Liu}, {Chen},
  {Cao}, {Xu}, {Li}, {Lin}, \& {Zhang}}]{ZhangTX2020}
{Zhang}, S.~N., {Tuo}, Y.~L., {Xiong}, S.~L., {et~al.} 2020, The Astronomer's
  Telegram, 13687, 1

\end{thebibliography}

\end{document}